# Effect of style of faulting on the probabilistic characterization of directionality of horizontal response spectral ordinates

Nathan Girmay[1*] and Eduardo Miranda[1]

[1]Blume Earthquake Engineering Center, Stanford University, Stanford, California, U.S.A.

Key points:

- We study and model the effect of faulting mechanism on directionality of horizontal response spectra.
- Directionality in strike-slip events differs notably from reverse-slip, especially at longer periods.
- Mechanism-specific models are developed to better estimate response spectra at specific orientations.

[*] Corresponding author: ngirmay3@stanford.edu

*Submitted to the Bulletin of the Seismological Society of America for review and possible publication*

# Abstract

Horizontal earthquake ground motions exhibit substantial changes in amplitude with changes in orientation. While the effects of style of faulting on central tendency measures of ground motion intensity such as RotD50 are well established in ground motion models, less attention has been given to how faulting mechanism influences the complete directional characteristics of ground motions. This study addresses this gap by conducting a comprehensive probabilistic characterization of ground motion directionality for strike-slip and reverse-slip earthquakes using 3,862 records with moment magnitudes greater than five. Two directionality parameters were examined across 58 oscillator periods ranging from 0.01 to 10 s. The directionality parameters examined consist of pseudo-acceleration response spectral ordinates at all non-redundant orientations normalized by either RotD100 or RotD50 intensities (denoted as $\eta$ and $\nu$, respectively). Nonlinear regression was used to develop mechanism-specific models for the geometric mean, logarithmic standard deviations, and full probability distributions of these parameters. Results show that strike-slip earthquakes exhibit notably different directionality compared to reverse-slip earthquakes, particularly at longer periods. Strike-slip earthquakes are, on average, 1% to 44% more polarized than reverse-slip earthquakes at periods above 0.5 s, and have RotD100/RotD50 ratios that are 1% to 3.7% larger. The developed models enable engineers and engineering seismologists to obtain mechanism-specific estimates of pseudo-acceleration response spectral ordinates at specific orientations, for applications such as regional seismic risk assessments and orientation-specific probabilistic seismic hazard analyses.

## Introduction

Earthquake ground motions at a site are fundamentally vector quantities, often recorded in two orthogonal horizontal components and one vertical component. The horizontal components of the ground motion are typically of greater concern to engineers because buildings are principally designed to withstand vertical gravity loads, making them comparatively more vulnerable to lateral shaking. Often, the 5%-damped pseudo-acceleration response spectral ordinate, henceforth referred to as the spectral ordinate, is used to represent the intensity in horizontal components. These horizontal spectral ordinates exhibit considerable variations in amplitude with changes in azimuth, in what is called directionality (Hong and Goda, 2007; Shahi and Baker, 2014). For instance, depending on oscillator frequency, spectral ordinates along the direction of maximum response are typically about 1.3 to 2 times larger than those in the orthogonal orientation (Hong and Goda, 2007; Poulos and Miranda, 2022). At a regional level, after the variability in median spectral ordinates, directionality has the second largest contribution to the variance of total losses for a group of buildings that share principal orientations (such as a regular city grid), contributing over 2.5 times as much to loss variability as the other sources of uncertainty (Poulos and Miranda, 2025). Therefore, ground motion directionality is significant, and its appropriate quantification is warranted to improve estimates of the shaking structures are likely to experience in future earthquakes.

Although the azimuth dependence of spectral ordinates has long been known, a single scalar value is often used to represent horizontal ground motion intensity in ground motion models (GMMs) and earthquake engineering. For example, most modern GMMs use the median intensity of all horizontal orientations (known as RotD50), whereas seismic design codes in the United States use the maximum intensity (known as RotD100). As such, most existing studies on ground motion directionality have focused on defining scalar measures of intensity (Boore et al., 2006;

Boore, 2010), or defining empirical relations between different scalar definitions of horizontal spectral ordinates (Beyer and Bommer, 2006; Watson-Lamprey and Boore, 2007; Shahi and Baker, 2014; Boore and Kishida, 2016; Poulos and Miranda, 2021). There has been less effort to quantify spectral amplitudes at arbitrary or specific azimuths and their variation with changes in orientation. In the first broad effort to characterize ground motion directionality, Hong and Goda (2007) used 592 records to investigate how spectral ordinates systematically decrease as one rotates away from the RotD100 orientation, which they defined as the major response axis. They observed that spectral ordinates, when normalized by the RotD100 intensity, tend to exhibit an approximately elliptical variation as one rotates away from the major response axis. Based on this observation, they developed a model of the variation in the mean of the normalized spectral intensities as a function of the angle from the major response axis for oscillators with periods from 0.2 to 3 s, and fitted probability distributions to the normalized intensity along the minor axis (i.e., the axis perpendicular to the major axis). Shahi and Baker (2014), using a similar approach but a much larger database of 3000 records, quantified how spectral intensities normalized by RotD50 vary with the angle from which RotD100 occurs, thereby enabling estimation of median intensities at arbitrary orientations using modern RotD50-based GMMs. However, they did not fit distributions to the normalized intensities at different angles of rotation. Poulos and Miranda (2022) extended the work of Hong and Goda (2007) and Shahi and Baker (2014) to provide the most comprehensive characterization of directionality to date. In their study, they used 5065 ground motion records to investigate the mean variation of spectral ordinates normalized by both RotD100 and RotD50 as a function of angle from RotD100, and fitted probability distributions and developed theoretical bounds for the ratios at all non-redundant orientations for 21 oscillator periods between 0.01 s and 10 s.

Although comprehensive, Poulos and Miranda (2022) (and preceding studies on directionality) aggregated records from different tectonic regimes and did not examine the effect of the style of faulting, which is known to have a notable influence on amplitudes of spectral ordinates. While the effects of magnitude, distance to the source, and local site conditions on spectral ordinates have been recognized for over 50 years, the effect of style of faulting has taken longer to be recognized and incorporated into GMMs. One of the first studies to investigate the effects of style of faulting was Aptikaev and Kopnichev (1980), which examined records from 59 earthquakes and concluded that the accuracy of strong motion prediction could be improved by considering the faulting mechanism. However, early subsequent studies provided contradictory results. For example, Crouse et al. (1988) separated data by faulting mechanism in subduction zones and found no significant differences in average pseudo-velocity spectral ordinates among thrust, normal, and strike-slip earthquakes. Meanwhile, Spudich et al. (1999) found that strike-slip faults may produce ground motions slightly larger than those of normal faults in extensional tectonic regimes. An early GMM that incorporated the effect of faulting style on the estimation of spectral ordinates was Boore et al. (1994), which showed discernible differences between ground motions from strike-slip and reverse-slip earthquakes, with median spectral ordinates from reverse-slip earthquakes up to 55% higher than those from strike-slip earthquakes for short periods, and with higher spectral ordinates for strike-slip earthquakes for periods longer than 1.5 s.

While most GMMs from the late 1990s in North America or New Zealand incorporated the effect of style of faulting on the geometric mean of response spectral ordinates (e.g., Abrahamson and Silva, 1997; Boore et al., 1997; Campbell, 1997; Sadigh et al., 1997; Zhao et al., 1997), the same was not true in European GMMs (e.g., Theodulidis and Papazachos, 1994; Garcia-Fernandez and Canas, 1995; Ambraseys et al., 1996; Sabetta and Pugliese, 1996). One of the first European models to incorporate the style of faulting into mechanism-independent GMMs was developed by Bommer et al. (2003) using the ground motion database assembled by Ambraseys et al. (1996). Most recent GMMs in North America (e.g., Abrahamson et al., 2014; Boore et al., 2014; Campbell and Bozorgnia, 2014; Chiou and Youngs, 2014; Gregor et al., 2014; Idriss, 2014; Goulet et al., 2021), in Japan (e.g., Zhao et al., 2016; Sung et al., 2024), in Europe (e.g., Bindi et al., 2011; Akkar et al., 2014; Douglas et al., 2014; Cauzzi et al., 2015; Kotha et al., 2020) and New Zealand (e.g., Lee et al., 2023) now incorporate the effect of style of faulting on the central tendency of spectral ordinates (typically RotD50).

Nonetheless, an aspect that has received much less attention is the effect of style of faulting on the directionality of spectral ordinates (i.e., looking beyond the effects of faulting mechanism on measures of central tendencies like RotD50). Hong and Goda (2010) conducted the first study of the effect of style of faulting on the directionality of spectral ordinates. They found that the directionality of records from in-slab earthquakes in Mexico was 6% and 18% higher than that of records from interplate earthquakes in Mexico for 0.2 s and 1.0 s oscillators, respectively, and that the directionality of California records was between the two. However, they did not study the orientation of maximum spectral ordinate nor the effect of style of faulting on this orientation. The first study to investigate the effect of style of faulting on the orientation of maximum spectral ordinates was conducted by Poulos and Miranda (2023a), in which they examined the probability distributions of the angular difference between the RotD100 orientation and the epicentral transverse orientation. Using records from the NGA-West2 database (Ancheta et al., 2014), they concluded that faulting mechanism has a significant influence on RotD100 orientation, with RotD100 orientations occurring close to the transverse for strike-slip earthquakes, but occurring at random orientations with respect to the transverse orientation for reverse-slip earthquakes.

Based on this observation, they developed a model for strike-slip earthquakes that can be used with RotD50-based GMMs to obtain spectral ordinates in specific orientations relative to the transverse orientation (Poulos and Miranda, 2023b). Nevertheless, as with most of the previously mentioned directionality studies, Poulos and Miranda did not examine the effects of the faulting mechanism on the complete directionality of spectral ordinates, despite their observations suggesting that directionality may be influenced by the tectonic regime.

Therefore, the objective of this study is to extend the work of Hong and Goda (2007), Shahi and Baker (2014), and Poulos and Miranda (2022), and characterize the effect of style of faulting on the directionality of horizontal earthquake ground motions. For this purpose, a set of 3862 ground motion records from the NGA-West2 database is used. The records are grouped by faulting mechanism, and for each mechanism, we perform a thorough probabilistic characterization of two main directionality parameters, which consist of 5% damped pseudo-acceleration spectral ordinates at all non-redundant orientations from the RotD100 orientation normalized by either the RotD100 or RotD50 intensities. These directionality parameters are quantified for 58 oscillator periods ranging from 0.01 to 10 s, and compared between strike-slip and reverse-slip earthquakes. Lastly, nonlinear regression is used to fit style-of-faulting specific models to the geometric mean, logarithmic standard deviations, and full probability distributions of the two directionality parameters.

## Parameters describing ground motion directionality

Although ground motions are typically recorded in two orthogonal components at a given site, response spectral ordinates in any orientation can be computed. This is achieved by subjecting a 5%-damped linear-elastic oscillator to the two orthogonal components, then combining the displacement responses and rotating over all nonredundant orientations. The orientation from which one starts rotating can differ depending on the application of interest. For example, Hong and Goda (2007) rotated ground motions starting from the major response axis (i.e., the orientation in which RotD100 occurs), and studied the pseudo-acceleration spectral ordinates at an angle $\phi$ from the RotD100 orientation (denoted $Sa(\phi)$). Studying spectral ordinates as a function of angle from RotD100 has the advantage that amplitudes tend to decrease as one rotates away from the major axis, allowing for a clearer comparison of total directionality across different records. To examine the directionality of different records, they proposed the parameter $\eta(\phi)$, which normalizes $Sa(\phi)$ by the maximum (i.e., RotD100) as follows

$$\eta(\phi) = \frac{Sa(\phi)}{Sa_{RotD100}} \tag{1}$$

The $\eta$ parameter provides a convenient normalization because its maximum value is bounded by 1 and its minimum by $\cos(\phi)$ (i.e., corresponding to the case of a fully linearly polarized ground motion). Additionally, if quantified a priori, $\eta$ can be used to obtain realizations of spectral ordinates at specific orientations, given a GMM for RotD100 intensity. A particularly useful special case is $\eta(90°)$, which serves as a metric for the level of polarization of a ground motion record (or a set) (e.g., as used in Girmay et al. (2024, 2025)).

While a useful metric for quantifying directionality, $\eta$ cannot be applied easily to seismic hazard analysis at present, since, to the best of our knowledge, no GMMs exist for RotD100 intensity. Instead, as mentioned in the introduction, most modern GMMs use the RotD50 intensity (e.g., Abrahamson et al., 2014; Boore et al., 2014; Campbell and Bozorgnia, 2014; Chiou and Youngs, 2014). Therefore, a second, more applicable directionality parameter that can be readily used with modern GMMs is the parameter $\nu(\phi)$, which normalizes $Sa(\phi)$ by the RotD50 intensity as follows

$$\nu(\phi) = \frac{Sa(\phi)}{Sa_{RotD50}} \tag{2}$$

In its original form, $\nu$ was first studied by Shahi and Baker (2014), and more thoroughly by Poulos and Miranda (2022). Others, such as Boore and Kishida (2016), have also studied the special case of $\nu(0°)$, which corresponds to the RotD100/RotD50 ratio used to convert seismic hazard curves used by the United States Geological Survey (USGS) to those used in U.S. building codes. The bounds for $\nu(\phi)$ are more nuanced than those for $\eta(\phi)$ and have been fully derived by Poulos and Miranda (2022).

Both directionality parameters ($\eta(\phi)$ and $\nu(\phi)$) exhibit significant record-to-record variability and depend on oscillator period. A more complete understanding of how faulting mechanism affects directionality can be developed by quantifying the mean, variability, and probability distributions of these two parameters across periods for records aggregated by style of faulting.

## Selected ground motions

We examine the effects of faulting mechanism on $\eta$ and $\nu$ using a subset of ground motion records from the NGA-West2 database (Ancheta et al., 2014). Since engineers are mostly interested in ground motions that can cause damage, we only consider events with moment magnitudes of five or greater. The chosen events are classified as either strike-slip (SS) or reverse-slip (RS) using the rake angle per the classification system used for the NGA-West2 database (i.e., rake angles within 30° of the perfect mechanism slip vector). Normal-slip earthquakes are not considered in this study because they were few in number, preventing us from obtaining statistically significant results. All records used in this study are from acceptably free-field sites per Boore et al. (2014) and are classified as NEHRP site classes B, C, or D. The latter filtering was used to ensure that we quantify directionality at sites that are not affected by strong site effects.

Lastly, to ensure that each record has a satisfactory signal-to-noise ratio across a wide range of periods (particularly at long periods), only records with a maximum PGV greater than 0.1 cm/s are used. Overall, this results in a total of 3862 ground motion records, of which 2220 are from RS events, and 1642 are from SS events.

## Effect of style-of-faulting on directionality parameters

Using the set of ground motions described above, $\eta$ and $\nu$ were computed for each record at rotation angles $\phi \in [-90°, 90°]$ with 1° increments across 58 oscillator periods ranging from 0.01 s to 10 s. Then, at each period, the natural logarithm of each parameter (i.e., $\ln \eta(\phi)$ or $\ln \nu(\phi)$) was calculated, and its statistics were computed after aggregating by style of faulting. The statistics are based on the natural logarithms of the directionality parameters since spectral accelerations follow a lognormal distribution fairly well. Figure 1 shows the geometric mean of $\eta$, denoted as $\mu_\eta$, at six periods as a function of the angle $\phi$, with solid lines indicating results for SS earthquakes and dashed lines indicating RS. Figure 2 is similar to Figure 1, but instead shows the geometric mean of $\nu$, denoted as $\mu_\nu$, at the same six periods for both SS and RS earthquakes. As described in Poulos and Miranda (2022), the geometric mean of $\eta$ and $\nu$ are directly related to the mean of the natural logarithm of each parameter as

$$\mu_\eta(\phi) = \exp(\mathrm{E}\,[\ln \eta(\phi)]) \quad (3)$$

and

$$\mu_\nu(\phi) = \exp(\mathrm{E}\,[\ln \nu(\phi)]) \quad (4)$$

where E[ ] is the expected value. Figures 1 and 2 indicate that as one rotates away from the orientation where the RotD100 intensity occurs (i.e., major response axis), $\mu_\eta$ and $\mu_\nu$ for both SS and RS events monotonically decrease. Regardless of the faulting mechanism, the rate of decrease of $\mu_\eta$ and $\mu_\nu$ is greater at longer oscillator periods, indicating that the level of polarization increases with period. At short periods, SS and RS earthquakes tend to have nearly identical $\mu_\eta$ and $\mu_\nu$, suggesting that style of faulting does not have a significant effect on directionality in this spectral region. However, as oscillator period increases, $\mu_\eta$ and $\mu_\nu$ for SS earthquakes significantly deviate from those for RS earthquakes. In particular, $\mu_\eta$ for SS earthquakes exhibits a notably larger decrease as one rotates away from the major response axis, with significantly lower values of $\mu_\eta(90°)$ when compared to RS earthquakes. This means that SS earthquakes tend to produce ground motions that are, on average, significantly more polarized than those from RS earthquakes. Similar observations hold in Figure 2 for $\mu_\nu$, with ground motions from SS earthquakes having $\mu_\nu(0°)$ values larger than and $\mu_\nu(90°)$ values smaller than those from RS earthquakes for periods longer than 1s. Since $\mu_\nu(0°)$ corresponds to the RotD100/RotD50 amplification factor often used to convert USGS seismic hazard curves to the RotD100 intensities used in building codes, an important implication of this observation is that style of faulting may need to be explicitly accounted for in these conversion factors.

In addition to quantifying the mean of $\eta$ and $\nu$, it is also very important to quantify the standard deviation of their natural logarithms, denoted $\sigma_{\ln\eta}$ and $\sigma_{\ln\nu}$ respectively. Figure 3 presents $\sigma_{\ln\eta}$ as a function of the angle $\phi$ for six different oscillator periods, while Figure 4 presents the same result but for $\sigma_{\ln\nu}$. What Figure 3 indicates is that, regardless of the faulting mechanism, $\sigma_{\ln\eta}$ tends to exhibit a smooth, monotonically increasing behavior as one rotates away from the major response axis. Similar to the geometric means, at short periods, there is no significant difference between $\sigma_{\ln\eta}$ for ground motions recorded in SS earthquakes and those in RS earthquakes. However, as the oscillator period increases, $\sigma_{\ln\eta}$ for SS events systematically deviate from those of RS, becoming larger, particularly at orientations beyond 60° from the major response axis. This means that spectral ordinates from SS events, on average, exhibit significantly larger variability when compared to those from RS, especially at orientations perpendicular to the major response axis. In contrast to $\sigma_{\ln\eta}$, $\sigma_{\ln\nu}$ does not exhibit a monotonic increase as one rotates away from $\phi = 0°$. Instead, Figure 4 indicates that $\sigma_{\ln\nu}$ initially decreases as one rotates away from the major response axis, reaching a minimum at approximately $\phi = \pm 45°$, and then increases beyond this rotation angle, reaching maximum variability in the orientation orthogonal to the major response axis. Evidently, for rotation angles near the major response axis, $\sigma_{\ln\nu}$ for SS and RS tend to be very similar regardless of period since the level of variability in these angles is small (i.e., less than 0.1). However, beyond $\phi = \pm 45°$, $\sigma_{\ln\nu}$ for SS becomes larger than that for RS, particularly at longer periods. Beyond $\sigma_{\ln\eta}$ and $\sigma_{\ln\nu}$, we also found that the total variability of spectral ordinates with orientation, referred to as component-to-component variability, $\sigma_{c2c}$, (Boore et al., 1997; Girmay and Miranda, 2026), for SS earthquakes tends to be notably larger than that for RS at longer periods (see Figure S1 and S2 in the electronic supplement). All of this indicates that, compared to RS earthquakes, SS earthquakes not only cause greater polarization in spectral responses but also exhibit greater variability in amplitude with changes in orientation.

While the differences in $\mu_\eta$ and $\mu_\nu$ between SS and RS events were compared in Figures 1 and 2, Figure 5 further quantifies these differences numerically. In particular, Figure 5(a) presents the ratio of $\mu_\eta$ computed from ground motions recorded in SS events to those in RS events (i.e., $\frac{\mu_{\eta,SS}}{\mu_{\eta,RS}}$) as a function of rotation angle $\phi$ at six oscillator periods. Within rotation angles approximately $\pm 30°$ from the major response axis, $\mu_\eta$ for SS and RS are approximately equal regardless of oscillator period. At orientations far from RotD100 and for most longer periods, SS earthquakes have $\mu_\eta$ values that are lower than those from RS earthquakes. Specifically, at 1 s, $\frac{\mu_{\eta,SS}(90°)}{\mu_{\eta,RS}(90°)}$ is equal to 0.99, and this goes down to 0.69 at 10 s, meaning that RS earthquakes are between 1% to 31% less polarized than SS earthquakes for periods longer than 1 s. In other words,

SS earthquakes are, on average, between 1% to 44% more polarized than RS earthquakes, which is a significant difference. While this has been suspected before, this is, to the best of the authors' knowledge, the first time that the differences in polarization of spectral ordinates between SS and RS have been numerically quantified. Interestingly, while not explicitly shown in Figure 5(a) (due to space limitations), it was found that SS earthquakes remain more polarized than RS earthquakes for oscillator periods as low as 0.5 s, but a reversal occurs below 0.5 s, where RS earthquakes become, on average, up to 1% more polarized than SS earthquakes.

Figure 5(b) is like Figure 5(a) but now quantifies the ratio of $\mu_\nu$ from SS events to that from RS events (i.e., $\frac{\mu_{\nu,SS}}{\mu_{\nu,RS}}$). As previously observed in Figure 2, there are no significant differences in $\mu_\nu$ between SS and RS events at periods less than 1s, with $\frac{\mu_{\nu,SS}}{\mu_{\nu,RS}}$ remaining close to unity at all rotation angles. However, at 5 s, SS earthquakes have $\mu_\nu\,(0°)$ values that are 1.1% larger than those for RS, increasing to 3.7% larger at 10 s. Since $\mu_\nu(0°)$ corresponds to the RotD100/RotD50 amplification factor, the difference in its value between SS and RS events suggests that mechanism-specific conversion factors may need to be considered in future seismic hazard applications. A similar comparison of $\sigma_{\ln\eta}$ and $\sigma_{\ln\nu}$ for SS and RS events is presented in Figure 6, with $\frac{\sigma_{\ln\eta_{SS}}}{\sigma_{\ln\eta_{RS}}}$ shown in Figure 6(a) and $\frac{\sigma_{\ln\nu_{SS}}}{\sigma_{\ln\nu_{RS}}}$ in Figure 6(b). The ratio $\frac{\sigma_{\ln\eta_{SS}}}{\sigma_{\ln\eta_{RS}}}$ becomes undefined at or near $\phi = 0°$ since, from the definition of the directionality parameter $\eta$, $\sigma_{\ln\eta}$ is zero along the major response axis regardless of the period or faulting mechanism. For rotation angles roughly within 60° of the major response axis, SS events tend to have lower $\sigma_{\ln\eta}$ and $\sigma_{\ln\nu}$ at most periods. At orientations perpendicular to RotD100, SS events have $\sigma_{\ln\eta}$ that get up to 34% greater and $\sigma_{\ln\nu}$ up to 41% larger than those for RS earthquakes, depending on oscillator period.

## Style of faulting dependent models of $\eta$ and $\nu$

Based on the results presented thus far, it is clear that the statistics of $\eta$ and $\nu$ differ between SS and RS earthquakes. Existing models for these parameters (e.g., Shahi and Baker, 2014; Poulos and Miranda, 2022) were developed using aggregated datasets that mix faulting mechanisms from crustal earthquakes, thereby averaging out mechanism-specific directionality effects. Therefore, new models that are specific to the style of faulting are required to capture differences in the statistics of $\eta$ and $\nu$ between SS and RS earthquakes. The empirical data presented in Figures 1 and 2 were used to develop mechanism-specific models for the geometric mean of $\eta$ and $\nu$ using the functional form given by

$$\hat{\mu}(\phi) = c_1 + c_2 \cos c_4 \, \phi + c_3 \cos \frac{9}{2} \phi \tag{5}$$

where $\hat{\mu}$ is the geometric mean of $\eta$ or $\nu$, $\phi$ is the rotation angle from RotD100 orientation, and coefficients $c_1 - c_4$ were estimated separately for SS and RS earthquakes at each oscillator period using nonlinear regression analysis. The fitted coefficients are provided in the electronic supplement of this article. Figures 7 and 8 use dashed lines to compare the resulting models of $\mu_\eta$ and $\mu_\nu$, respectively, to statistical results shown in continuous lines at six oscillator periods for SS and RS earthquakes. As shown in these figures, the models fit the statistical results very well across all periods for both faulting mechanisms, capturing the monotonic decrease in the geometric mean as one rotates away from the major response axis.

A model was also developed for $\sigma_{\ln \eta}$ and $\sigma_{\ln \nu}$ using the statistical results presented in Figures 3 and 4. The model for the logarithmic standard deviation of $\eta$ or $\nu$ takes the functional form

$$\hat{\sigma}_{\ln}(\phi) = \frac{c_1 + c_2 \cos c_6 \, \phi + c_3 \cos 4\phi}{1 + c_4 \cos c_6 \, \phi + c_5 \cos 4\phi} \tag{6}$$

where $\hat{\sigma}_{\ln}$ is the logarithmic standard deviation of $\eta$ or $\nu$. Again, the coefficients $c_1 - c_6$ were estimated separately for SS and RS earthquakes at 58 oscillator periods, and their values are provided in the electronic supplement. Figures 9 and 10 compare the new models for $\sigma_{\ln \eta}$ and $\sigma_{\ln \nu}$ with statistical results across oscillator periods, showing a good fit.

The resulting models for the geometric means and logarithmic standard deviations of $\eta$ or $\nu$ can be compared with existing directionality models of crustal earthquakes to assess the consequences of aggregating different faulting styles. To this end, Figure 11 compares the models for $\mu_\eta$ and $\mu_\nu$ developed in this paper (equation (5)) with the models from Shahi and Baker (2014) (henceforth SB14) and Poulos and Miranda (2022) (henceforth PM22). Note that SB14 did not model $\mu_\eta$ but provided numerical estimates for $\mu_\nu$ at 10° rotation angle increments for 21 periods ranging from 0.01 s to 10. Therefore, the model for $\mu_\eta$ is compared only with PM22, while the $\mu_\nu$ model is compared with both SB14 and PM22. Evidently, at very short periods, where faulting mechanism has little influence on directionality, all models provide estimates of $\mu_\eta$ and $\mu_\nu$ that are very similar. At 5 s, despite all models being developed using different datasets, they still generally provide similar estimates of $\mu_\eta$ and $\mu_\nu$ for $\phi$ angles below 60°. However, for $\phi > 60°$, SB14 results in $\mu_\nu$ close to the SS model developed in this study, while PM22 falls right in between the SS and RS models. The differences between SB14 and PM22, despite both groups aggregating different faulting styles, may be attributed to their dataset. For example, if SB14 considered numerous small-magnitude events in the NGA-West2 database, the dataset would be skewed towards more SS events, since approximately 70% of the records in the NGA-West2 are from (mainly) California, which is dominated by SS faults. Meanwhile, PM22 only considered magnitude equal or larger than five. At very long periods such as 10 s, both SB14 and PM22 estimates for $\mu_\eta$ and $\mu_\nu$ fall in between the mechanism-specific models developed in this paper, with both models overestimating $\mu_\nu(90°)$ for SS events by over 22%. Consequently, PM22 overestimates $\mu_\eta(90°)$ for SS events by 27% and underestimates $\mu_\eta(90°)$ by 15% for RS events. Similarly, Figure 12 compares the models for $\sigma_{\ln\eta}$ and $\sigma_{\ln\nu}$ (equation (6)) with PM22 at different oscillator periods. Much like that observed for the geometric means, the $\sigma_{\ln\eta}$ and $\sigma_{\ln\nu}$ estimated by PM22 at short periods are nearly identical to the mechanism-specific models developed herein. However, at longer periods, PM22 estimates for $\sigma_{\ln\eta}$ and $\sigma_{\ln\nu}$ tend to fall between the mechanism-specific models, particularly at $\phi > 60°$, where PM22 underestimates the variability for SS events but overestimates for RS. Ultimately, Figures 11 and 12 clearly indicate that combining records from different tectonic regimes results in the averaging out of the mechanism-specific directionality effects.

In certain applications, such as for regional risk estimation (e.g., Bantis, Heresi, et al., 2025; Bantis, Miranda, et al., 2025), the complete probability distribution of $\eta$ or $\nu$ may be needed. Studies by Hong and Goda (2007) and more recently by Poulos and Miranda (2022) have shown that the probability density function of $\eta$ or $\nu$ can be represented using a four-parameter Beta distribution as follows:

$$f_x(x;\alpha,\beta,r,s) = \frac{(x-r)^{\alpha-1}(s-x)^{\beta-1}}{(s-r)^{\alpha+\beta-1}B(\alpha,\beta)} \tag{7}$$

where *x* is either $\eta$ or $\nu$, $\alpha$ and $\beta$ are shape parameters, *r* and *s* are supports, and *B( )* is the Beta function. The supports of the distribution at each rotation angle $\phi$ are period-independent and correspond to the lower and upper bounds of $\eta$ or $\nu$ derived by Poulos and Miranda (2022). Maximum likelihood estimation was used to estimate the shape parameters for the distributions of $\eta$ and $\nu$ for SS and RS earthquakes at each oscillator period and for $\phi \in [-90°, 90°]$ with 1° increment in rotation angle. The period-independent fitted shape parameters are provided in the electronic supplement. Figure 13 compares the empirical and fitted distributions of $\eta$ for RS and SS events at four rotation angles and for 5 oscillator periods. Figure 14 presents similar results for the distributions of $\nu$. As illustrated for these selected periods, the model density function fits the empirical distributions fairly well. Generally, the probability distributions of SS and RS are similar at rotation angles close to the major response axis, but differ at orientations far from the orientation of RotD100.

Beyond the statistical moments and probability distributions studied above, a comparison of correlations of $\eta$ or $\nu$ at different rotation angles were performed for SS and RS earthquakes. In general, the correlation of spectral ordinates at different orientations was found to be less sensitive to style of faulting, and thus mechanism-specific models were not developed (although some minor differences do exist). For this purpose, it may be justifiable to use the correlation model in Poulos and Miranda (2022) for both SS and RS events. The interested reader is referred to Figures S3 – S8 in the electronic supplement, which present the correlations. Ultimately, the models presented in this study can be combined with existing GMMs to generate more realistic estimates of spectral ordinates at specific orientations accounting for style of faulting, for use in earthquake engineering and probabilistic seismic hazard assessments.

## Summary and conclusions

This study extends previous work on ground motion directionality by examining the effect of style of faulting on the directional characteristics of earthquake ground motions. Using a subset of 3862 ground motion records from the NGA-West2 database, we conducted a comprehensive probabilistic characterization of two key directionality parameters for ground motions recorded in strike-slip (SS) and reverse-slip (RS) earthquakes across 58 oscillator periods ranging from 0.01 to 10 s. The directionality parameters examined consist of spectral ordinates at all non-redundant orientations normalized by either RotD100 or RotD50 intensities (denoted as $\eta$ and $\nu$, respectively).

Results show that ground motions from strike-slip earthquakes exhibit notably different directionality characteristics compared to those from reverse-slip earthquakes, particularly at periods longer than 1 s. Specifically, we found that strike-slip earthquakes produce ground motions that are between 1% and 44% more polarized than reverse-slip earthquakes at periods above 0.5 s, with the degree of polarization increasing with increasing oscillator period. Furthermore, strike-slip earthquakes produce ground motions that have RotD100/RotD50 ratios that are, on average, 1% to 3.7% larger than those for reverse-slip earthquakes at periods exceeding 1 s. At short periods (less than 1 s), the directionality characteristics of both faulting mechanisms are nearly identical, indicating that style of faulting does not significantly affect ground motion polarization in this period range. These findings have important practical implications for earthquake engineering and seismic hazard analysis. The observed differences in RotD100/RotD50 ratios between faulting mechanisms suggest that mechanism-specific conversion factors should be used when converting USGS seismic hazard curves to the RotD100 intensities commonly used in U.S. building codes. The directionality parameters characterized in this study can be used in conjunction with existing ground motion models to obtain realizations of spectral ordinates at specific orientations, such as along a building's principal axes. This capability is particularly valuable for seismic hazard evaluations at sites where major response axes can be estimated in advance, an increasingly feasible task given ongoing research developments in this area.

Style-of-faulting dependent models were developed using nonlinear regression analyses to characterize the geometric mean, logarithmic standard deviations, and full probability distributions of both directionality parameters. These models provide earthquake engineers and seismic hazard analysts with tools to explicitly account for faulting mechanism effects on ground motion intensities at specific orientations, particularly for longer-period structures that are most sensitive to these differences.

It should be noted that this study did not examine normal-slip earthquakes due to the limited availability of ground motions from this faulting style in the available crustal ground motion dataset. For applications involving normal-faulting events, we recommend using existing directionality models that aggregate data from different tectonic regimes until sufficient data become available for mechanism-specific characterization. Ultimately, incorporating these mechanism-specific directionality models into probabilistic seismic hazard analysis frameworks represents an important next step toward improving the accuracy of orientation-dependent seismic hazard and risk assessments.

## Acknowledgements

The authors express their gratitude to all the agencies that install and maintain seismic instrumentation for collecting, processing, and distributing the ground motions used in this study. This investigation would not have been possible without this instrumentation and data.

## Funding

The authors acknowledge the Shah Fellowship on Catastrophic Risk from the Department of Civil and Environmental Engineering at Stanford University for providing funding to support the PhD studies of the first author. Partial funding for this study was provided by the California Strong Motion Instrumentation Program through agreement number 1023-003.

## Declaration of competing interests

The authors acknowledge there are no conflicts of interest.

## Data and resources

All ground motion records used in this study were sourced from the Next Generation Attenuation (NGA)-West2 ground motion database, which is developed and maintained by the Pacific Earthquake Engineering Research (PEER) center (available online at https://ngawest2.berkeley.edu/, last accessed March 2024). The parameters for the fitted models are available in the electronic supplement of this article.

## References

Abrahamson, N. A., and W. J. Silva, 1997, Empirical Response Spectral Attenuation Relations for Shallow Crustal Earthquakes, *Seismol. Res. Lett.*, 68, no. 1, 94–127, doi: 10.1785/gssrl.68.1.94.

Abrahamson, N. A., W. J. Silva, and R. Kamai, 2014, Summary of the ASK14 Ground Motion Relation for Active Crustal Regions, *Earthq. Spectra*, 30, no. 3, 1025–1055, doi: 10.1193/070913EQS198M.

Akkar, S., M. A. Sandıkkaya, and J. J. Bommer, 2014, Empirical ground-motion models for point- and extended-source crustal earthquake scenarios in Europe and the Middle East, *Bull. Earthq. Eng.*, 12, no. 1, 359–387, doi: 10.1007/s10518-013-9461-4.

Ambraseys, N. N., K. A. Simpson, and J. J. Bommer, 1996, Prediction of Horizontal Response Spectra in Europe, *Earthq. Eng. Struct. Dyn.*, 25, no. 4, 371–400, doi: 10.1002/(SICI)1096-9845(199604)25:4<371::AID-EQE550>3.0.CO;2-A.

Ancheta, T. D., R. B. Darragh, J. P. Stewart, E. Seyhan, W. J. Silva, B. S.-J. Chiou, K. E. Wooddell, R. W. Graves, A. R. Kottke, D. M. Boore, *et al.*, 2014, NGA-West2 Database, Earthq. Spectra, 30, no. 3, 989–1005, doi: 10.1193/070913EQS197M.

Aptikaev, N., and J. Kopnichev, 1980, Correlation between seismic vibration parameters and type of faulting, in *Proceedings of the Seventh World Conference on Earthquake Engineering* Istanbul, Turkey.

Bantis, J., Heresi, P., Poulos, A., & Miranda, E. (2025). Framework for regional seismic risk assessments of groups of tall buildings. *Earthq. Eng. Struct. Dyn.*, *54, no. 3, 833–850, doi: 10.1002/eqe.4283*.

Bantis, J., Miranda, E., & Heresi, P. (2025). Impact of urban layout and ground motion directionality on building responses and damages in San Francisco during the Loma Prieta earthquake. *Earthq. Spectra*, *41, no. 5, 3417–3446, doi: 10.1177/87552930251378230*.

Beyer, K., and J. J. Bommer, 2006, Relationships between Median Values and between Aleatory Variabilities for Different Definitions of the Horizontal Component of Motion, *Bull. Seismol. Soc. Am.*, 96, no. 4A, 1512–1522, doi: 10.1785/0120050210.

Bindi, D., F. Pacor, L. Luzi, R. Puglia, M. Massa, G. Ameri, and R. Paolucci, 2011, Ground motion prediction equations derived from the Italian strong motion database, *Bull. Earthq. Eng.*, 9, no. 6, 1899–1920, doi: 10.1007/s10518-011-9313-z.

Bommer, J. J., J. Douglas, and F. O. Strasser, 2003, Style-of-Faulting in Ground-Motion Prediction Equations, *Bull. Earthq. Eng.*, 1, no. 2, 171–203, doi: 10.1023/A:1026323123154.

Boore, D. M., W. B. Joyner, and T. E. Fumal, 1997, Equations for Estimating Horizontal Response Spectra and Peak Acceleration from Western North American Earthquakes: A Summary of Recent Work, *Seismol. Res. Lett.*, 68, no. 1, 128–153, doi: 10.1785/gssrl.68.1.128.

Boore, D. M., W. B. Joyner, and T. E. Fumal, 1994, Estimation of response spectra and peak accelerations from western north american earthquakes: an interim report - part 2, *Open-File Report 94–127*, U.S. Geological Survey, Menlo Park, California.

Boore, D. M., J. Watson-Lamprey, and N. A. Abrahamson, 2006, Orientation-Independent Measures of Ground Motion, *Bull. Seismol. Soc. Am.*, 96, no. 4A, 1502–1511, doi: 10.1785/0120050209.

Boore, D. M., 2010, Orientation-Independent, Nongeometric-Mean Measures of Seismic Intensity from Two Horizontal Components of Motion, *Bull. Seismol. Soc. Am.*, 100, no. 4, 1830–1835, doi: 10.1785/0120090400.

Boore, D. M., J. P. Stewart, E. Seyhan, and G. M. Atkinson, 2014, NGA-West2 Equations for Predicting PGA, PGV, and 5% Damped PSA for Shallow Crustal Earthquakes, *Earthq. Spectra*, 30, no. 3, 1057–1085, doi: 10.1193/070113EQS184M.

Boore, D. M., and T. Kishida, 2016, Relations between Some Horizontal-Component Ground-Motion Intensity Measures Used in Practice, *Bull. Seismol. Soc. Am.*, 107, no. 1, 334–343, doi: 10.1785/0120160250.

Campbell, K. W., 1997, Empirical Near-Source Attenuation Relationships for Horizontal and Vertical Components of Peak Ground Acceleration, Peak Ground Velocity, and Pseudo-Absolute Acceleration Response Spectra, *Seismol. Res. Lett.*, 68, no. 1, 154–179, doi: 10.1785/gssrl.68.1.154.

Campbell, K. W., and Y. Bozorgnia, 2014, NGA-West2 Ground Motion Model for the Average Horizontal Components of PGA, PGV, and 5% Damped Linear Acceleration Response Spectra, *Earthq. Spectra*, 30, no. 3, 1087–1115, doi: 10.1193/062913EQS175M.

Cauzzi, C., E. Faccioli, M. Vanini, and A. Bianchini, 2015, Updated predictive equations for broadband (0.01–10 s) horizontal response spectra and peak ground motions, based on a global dataset of digital acceleration records, *Bull. Earthq. Eng.*, 13, no. 6, 1587–1612, doi: 10.1007/s10518-014-9685-y.

Chiou, B. S.-J., and R. R. Youngs, 2014, Update of the Chiou and Youngs NGA Model for the Average Horizontal Component of Peak Ground Motion and Response Spectra, *Earthq. Spectra*, 30, no. 3, 1117–1153, doi: 10.1193/072813EQS219M.

Crouse, C. B., Y. K. Vyas, and B. A. Schell, 1988, Ground motions from subduction-zone earthquakes, *Bull. Seismol. Soc. Am.*, 78, no. 1, 1–25, doi: 10.1785/BSSA0780010001.

Douglas, J., S. Akkar, G. Ameri, P.-Y. Bard, D. Bindi, J. J. Bommer, S. S. Bora, F. Cotton, B. Derras, M. Hermkes, *et al.*, 2014, Comparisons among the five ground-motion models developed using RESORCE for the prediction of response spectral accelerations due to earthquakes in Europe and the Middle East, *Bull. Earthq. Eng.*, 12, no. 1, 341–358, doi: 10.1007/s10518-013-9522-8.

Garcia-Fernandez, M., and J. A. Canas, 1995, Regional peak ground acceleration estimates in the Iberian peninsula, in *Proceedings of the fifth international conference on seismic zonation* Nice, 1029–1034.

Girmay, N., A. Poulos, and E. Miranda, 2024, Directionality and polarization of response spectral ordinates in the 2023 Kahramanmaras, Türkiye earthquake doublet, *Earthq. Spectra*, 40, no. 1, 486–504, doi: 10.1177/87552930231203989.

Girmay, N., A. Poulos, and E. Miranda, 2025, Evaluation of directionality in physics-based ground motion simulations of strike-slip earthquakes, *Earthq. Spectra*, 41, no. 1, 436–456, doi: 10.1177/87552930241270555.

Girmay, N., and E. Miranda, 2026, Component-To-Component Variability of Response Spectral Ordinates from Strike-Slip Earthquakes, *Earthq. Spectra*, 42, no. 2, e70045, doi: 10.1002/esp4.70045.

Goulet, C. A., Y. Bozorgnia, N. Kuehn, L. Al Atik, R. R. Youngs, R. W. Graves, and G. M. Atkinson, 2021, NGA-East Ground-Motion Characterization model part I: Summary of products and model development, *Earthq. Spectra*, 37, no. 1_suppl, 1231–1282, doi: 10.1177/87552930211018723.

Gregor, N., N. A. Abrahamson, G. M. Atkinson, D. M. Boore, Y. Bozorgnia, K. W. Campbell, B. S.-J. Chiou, I. M. Idriss, R. Kamai, E. Seyhan, *et al.*, 2014, Comparison of NGA-West2 GMPEs, *Earthq. Spectra*, 30, no. 3, 1179–1197, doi: 10.1193/070113EQS186M.

Hong, H. P., and K. Goda, 2007, Orientation-Dependent Ground-Motion Measure for Seismic-Hazard Assessment, *Bull. Seismol. Soc. Am.*, 97, no. 5, 1525–1538, doi: 10.1785/0120060194.

Hong, H. P., and K. Goda, 2010, Characteristics of horizontal ground motion measures along principal directions, *Earthq. Eng. Eng. Vib.*, 9, no. 1, 9–22, doi: 10.1007/s11803-010-9048-x.

Idriss, I. M., 2014, An NGA-West2 Empirical Model for Estimating the Horizontal Spectral Values Generated by Shallow Crustal Earthquakes, *Earthq. Spectra*, 30, no. 3, 1155–1177, doi: 10.1193/070613EQS195M.

Kotha, S. R., G. Weatherill, D. Bindi, and F. Cotton, 2020, A regionally-adaptable ground-motion model for shallow crustal earthquakes in Europe, *Bull. Earthq. Eng.*, 18, no. 9, 4091–4125, doi: 10.1007/s10518-020-00869-1.

Lee, R. L., B. A. Bradley, E. F. Manea, J. A. Hutchinson, and S. S. Bora, 2023, Evaluation of Empirical Ground-Motion Models for the 2022 New Zealand National Seismic Hazard Model Revision, *Bull. Seismol. Soc. Am.*, 114, no. 1, 311–328, doi: 10.1785/0120230180.

Poulos, A., and E. Miranda, 2021, Relations between MaxRotD50 and Some Horizontal Components of Ground-Motion Intensity Used in Practice, *Bull. Seismol. Soc. Am.*, 111, no. 4, 2167–2176, doi: 10.1785/0120200364.

Poulos, A., and E. Miranda, 2022, Probabilistic characterization of the directionality of horizontal earthquake response spectra, *Earthq. Eng. Struct. Dyn.*, 51, no. 9, 2077–2090, doi: 10.1002/eqe.3654.

Poulos, A., and E. Miranda, 2023a, Effect of Style of Faulting on the Orientation of Maximum Horizontal Earthquake Response Spectra, *Bull. Seismol. Soc. Am.*, 113, no. 5, 2092–2105, doi: 10.1785/0120230001.

Poulos, A., and E. Miranda, 2023b, Modification of Ground-Motion Models to Estimate Orientation-Dependent Horizontal Response Spectra in Strike-Slip Earthquakes, *Bull. Seismol. Soc. Am.*, 113, no. 6, 2718–2729, doi: 10.1785/0120230084.

Poulos, A., and E. Miranda, 2025, Accounting for ground motion directionality and building orientations in urban seismic risk analysis, *Earthq. Spectra*, 41, no. 2, 1780–1800, doi: 10.1177/87552930251315751.

Sabetta, F., and A. Pugliese, 1996, Estimation of response spectra and simulation of nonstationary earthquake ground motions, *Bull. Seismol. Soc. Am.*, 86, no. 2, 337–352, doi: 10.1785/BSSA0860020337.

Sadigh, K., C.-Y. Chang, J. A. Egan, F. Makdisi, and R. R. Youngs, 1997, Attenuation Relationships for Shallow Crustal Earthquakes Based on California Strong Motion Data, *Seismol. Res. Lett.*, 68, no. 1, 180–189, doi: 10.1785/gssrl.68.1.180.

Shahi, S. K., and J. W. Baker, 2014, NGA-West2 Models for Ground Motion Directionality, *Earthq. Spectra*, 30, no. 3, 1285–1300, doi: 10.1193/040913EQS097M.

Spudich, P., W. B. Joyner, A. G. Lindh, D. M. Boore, B. M. Margaris, and J. B. Fletcher, 1999, SEA99: A revised ground motion prediction relation for use in extensional tectonic regimes, *Bull. Seismol. Soc. Am.*, 89, no. 5, 1156–1170, doi: 10.1785/BSSA0890051156.

Sung, C., H. Miyake, N. Abrahamson, and N. Morikawa, 2024, Nonergodic Ground-Motion Models for Subduction Zone and Crustal Earthquakes in Japan, *Bull. Seismol. Soc. Am.*, 114, no. 3, 1717–1738, doi: 10.1785/0120230258.

Theodulidis, N. P., and B. C. Papazachos, 1994, Dependence of strong ground motion on magnitude-distance, site geology and macroseismic intensity for shallow earthquakes in Greece: II, horizontal pseudovelocity, *Soil Dyn. Earthq. Eng.*, 13, no. 5, 317–343, doi: 10.1016/0267-7261(94)90024-8.

Watson-Lamprey, J. A., and D. M. Boore, 2007, Beyond SaGMRotI: Conversion to SaArb, Sasn, and SaMaxRot, *Bull. Seismol. Soc. Am.*, 97, no. 5, 1511–1524, doi: 10.1785/0120070007.

Zhao, J. X., D. J. Dowrick, and G. H. McVerry, 1997, Attenuation of peak ground accelerations in New Zealand earthquakes, *Bull. N. Z. Soc. Earthq. Eng.*, 30, no. 2, 133–158, doi: 10.5459/bnzsee.30.2.133-158.

Zhao, J. X., F. Jiang, P. Shi, H. Xing, H. Huang, R. Hou, Y. Zhang, P. Yu, X. Lan, D. A. Rhoades, *et al.*, 2016, Ground-Motion Prediction Equations for Subduction Slab Earthquakes in Japan Using Site Class and Simple Geometric Attenuation Functions, *Bull. Seismol. Soc. Am.*, 106, no. 4, 1535–1551, doi: 10.1785/0120150056.

## List of figure captions

| | |
|---|---|
| Figure 1: | Geometric mean of $\eta$, which represents pseudo-acceleration spectral ordinates at rotation angle $\phi$ from the major response axis (i.e., orientation where RotD100 occurs) normalized by the RotD100 intensity. Solid lines indicate results for ground motions recorded in strike-slip earthquakes, and dashed lines indicate results for ground motions recorded in reverse-slip earthquakes. |
| Figure 2: | Geometric mean of $\nu$, which represents pseudo-acceleration spectral ordinates at rotation angle $\phi$ from the major response axis normalized by the RotD50 intensity. Solid lines indicate results for ground motions recorded in strike-slip earthquakes, and dashed lines indicate results for ground motions recorded in reverse-slip earthquakes. |
| Figure 3: | Logarithmic standard deviations of $\eta$. Solid lines represent variabilities computed from ground motions recorded in strike-slip earthquakes, and dashed lines represent variabilities computed from ground motions recorded in reverse-slip earthquakes. |
| Figure 4: | Logarithmic standard deviations of $\nu$. Solid lines represent variabilities computed from ground motions recorded in strike-slip earthquakes, and dashed lines represent variabilities computed from ground motions recorded in reverse-slip earthquakes. |
| Figure 5: | Ratios of: (a) geometric mean of $\eta$ for ground motions recorded in strike-slip earthquakes to that from ground motions recorded in reverse-slip earthquakes; (b) geometric mean of $\nu$ for ground motions recorded in strike-slip earthquakes to that from ground motions recorded in reverse-slip earthquakes. |
| Figure 6: | Ratios of: (a) logarithmic standard deviations of $\eta$ for strike-slip ground motion records to those for reverse-slip records; (b) logarithmic standard deviations of $\nu$ for strike-slip ground motion records to those for reverse-slip records. |
| Figure 7: | Comparison of faulting mechanism specific statistical results and fitted models for the geometric mean of $\eta$. Solid lines indicate empirical results, and dashed lines represent the model (using equation 5). |
| Figure 8: | Comparison of faulting mechanism specific statistical results and fitted models for the geometric mean of $\nu$ . Solid lines indicate empirical results, and dashed lines represent the model (using equation 5). |
| Figure 9: | Comparison of faulting mechanism specific statistical results and fitted models for the logarithmic standard deviations of $\eta$. Solid lines indicate empirical results, and dashed lines represent the model (using equation 6). |
| Figure 10: | Comparison of faulting mechanism specific statistical results and fitted models for the logarithmic standard deviations of $\nu$. Solid lines indicate empirical results, and dashed lines represent the model (using equation 6). |
| Figure 11: | Comparison of faulting mechanism specific models for the geometric mean of $\eta$ and $\nu$ (equation 5) from this study with models by Shahi and Baker (2014) and |
| | Poulos and Miranda (2022) for crustal earthquakes. The top row presents the comparison for $\eta$ while the bottom shows the comparison for $\nu$. |
| Figure 12: | Comparison of faulting mechanism-specific models for the logarithmic standard deviations of $\eta$ and $\nu$ developed in this study (equation 6) with the model by Poulos and Miranda (2022). The top row presents the comparison for $\eta$ while the bottom shows the comparison for $\nu$. |
| Figure 13: | Empirical and fitted probability distributions of $\eta$ for strike-slip earthquakes (shown in grey) and reverse-slip earthquakes (shown in red) at four rotation angles and five oscillator periods. |
| Figure 14: | Empirical and fitted probability distributions of $\nu$ for strike-slip earthquakes (shown in grey) and reverse-slip earthquakes (shown in red) at four rotation angles and five oscillator periods. |

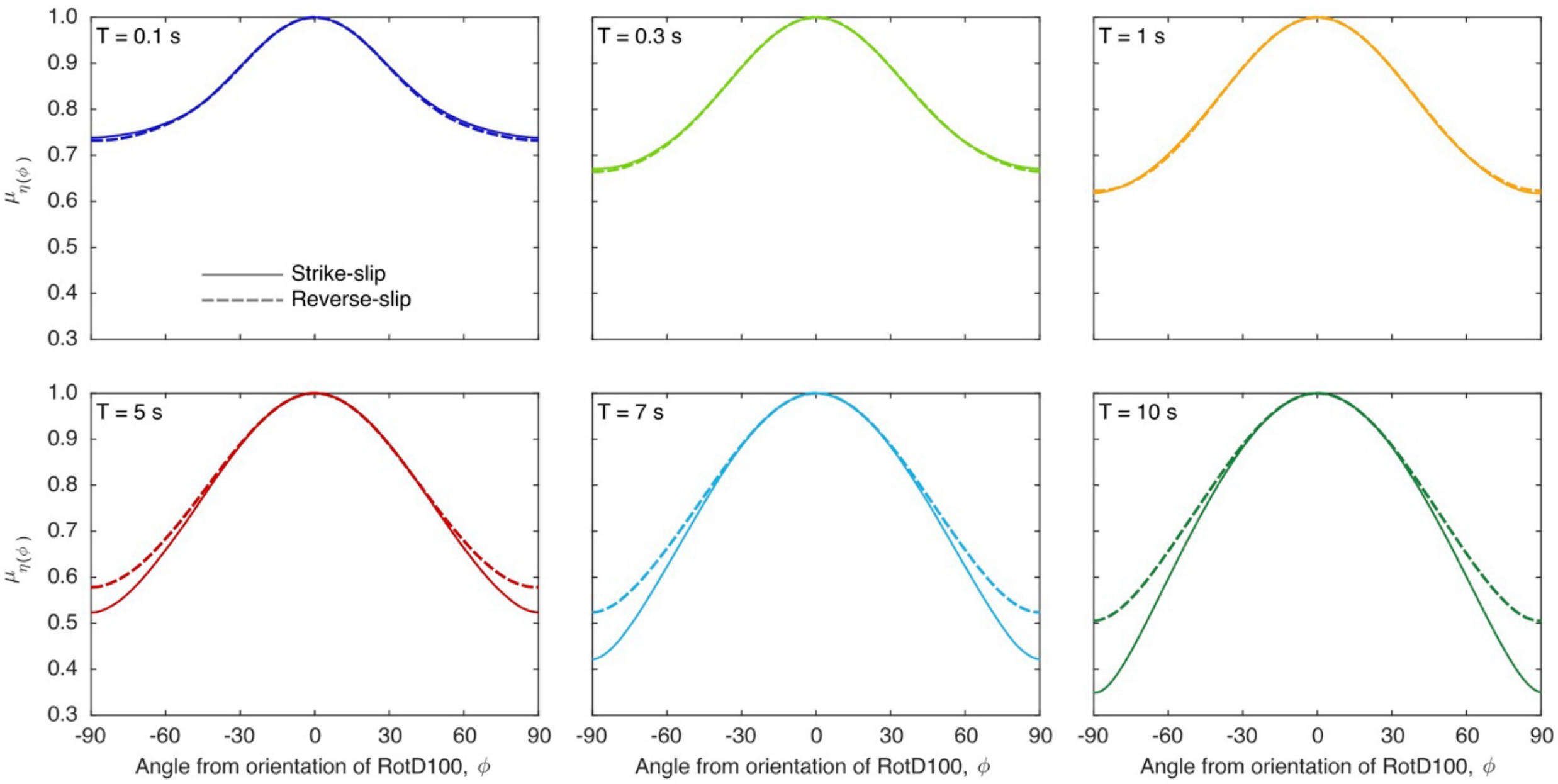

**Figure 1:** Geometric mean of $\eta$, which represents pseudo-acceleration spectral ordinates at rotation angle $\phi$ from the major response axis (i.e., orientation where RotD100 occurs) normalized by the RotD100 intensity. Solid lines indicate results for ground motions recorded in strike-slip earthquakes, and dashed lines indicate results for ground motions recorded in reverse-slip earthquakes.

***Alt-text:*** *Line graphs showing the geometric mean of spectral accelerations at given orientation from RotD100 normalized by RotD100, indicating that geometric means decrease as one rotates away from the major response axis.*

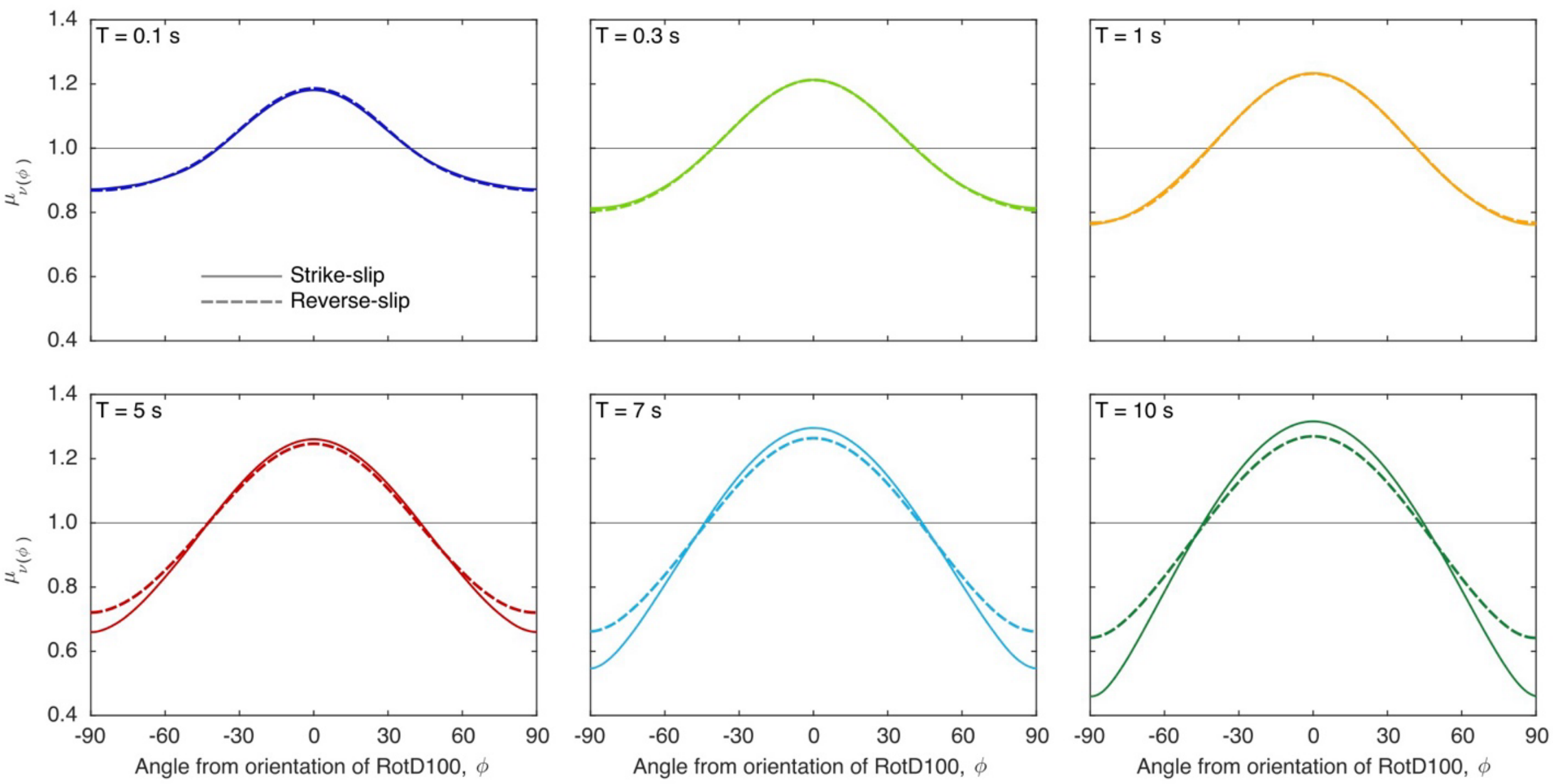

**Figure 2:** Geometric mean of $\nu$, which represents pseudo-acceleration spectral ordinates at rotation angle $\phi$ from the major response axis normalized by the RotD50 intensity. Solid lines indicate results for ground motions recorded in strike-slip earthquakes, and dashed lines indicate results for ground motions recorded in reverse-slip earthquakes.

***Alt-text:*** *Line graphs showing the geometric mean of spectral accelerations at given orientation from RotD100 normalized by RotD50, indicating that geometric means decrease as one rotates away from the major response axis.*

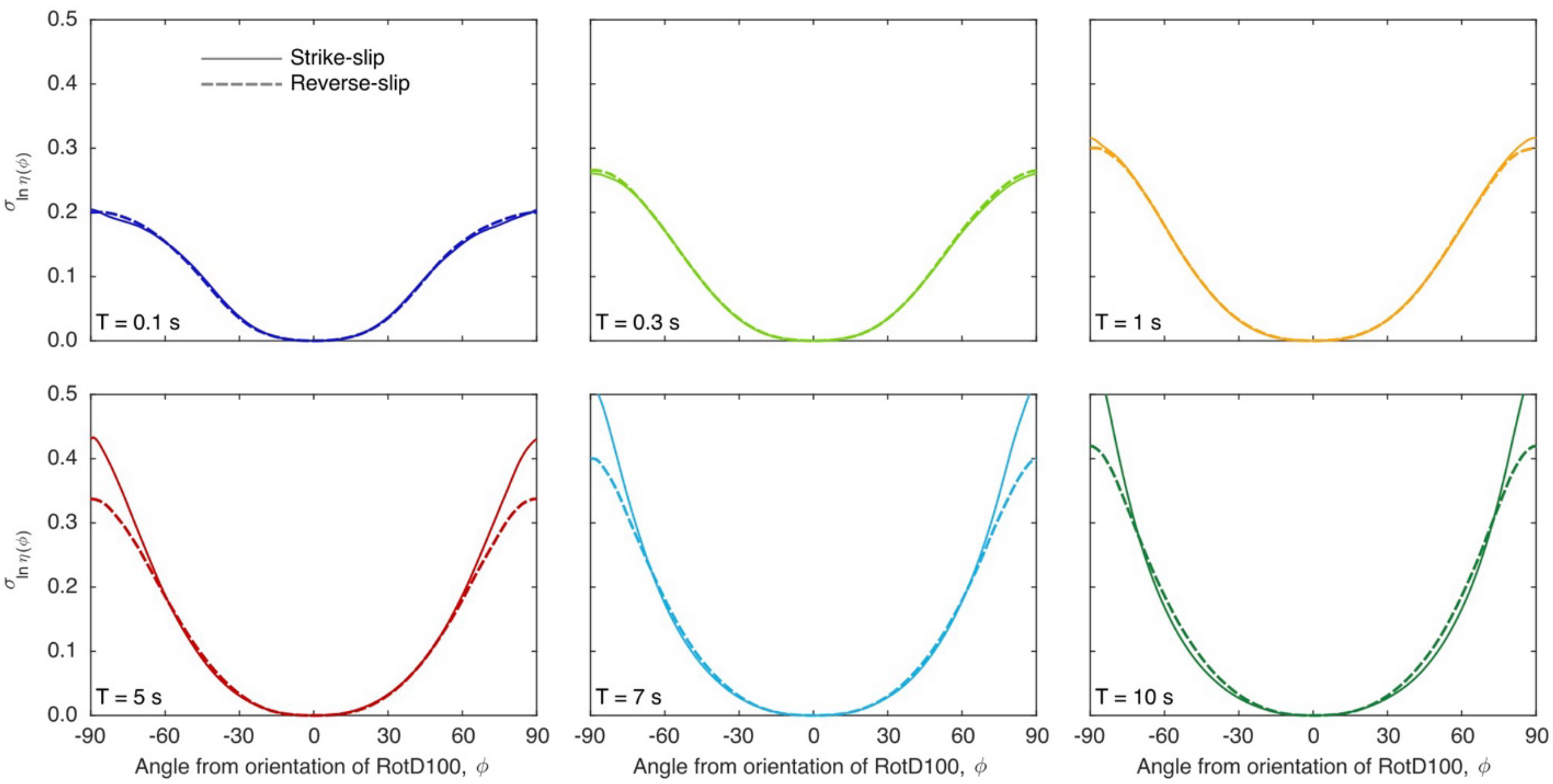

**Figure 3:** Logarithmic standard deviations of $\eta$. Solid lines represent variabilities computed from ground motions recorded in strike-slip earthquakes, and dashed lines represent variabilities computed from ground motions recorded in reverse-slip earthquakes.

***Alt-text:*** *Line graphs showing the logarithmic standard deviations of spectral accelerations at given orientation from RotD100 normalized by RotD100, indicating that standard deviations increase as one rotates away from the major response axis.*

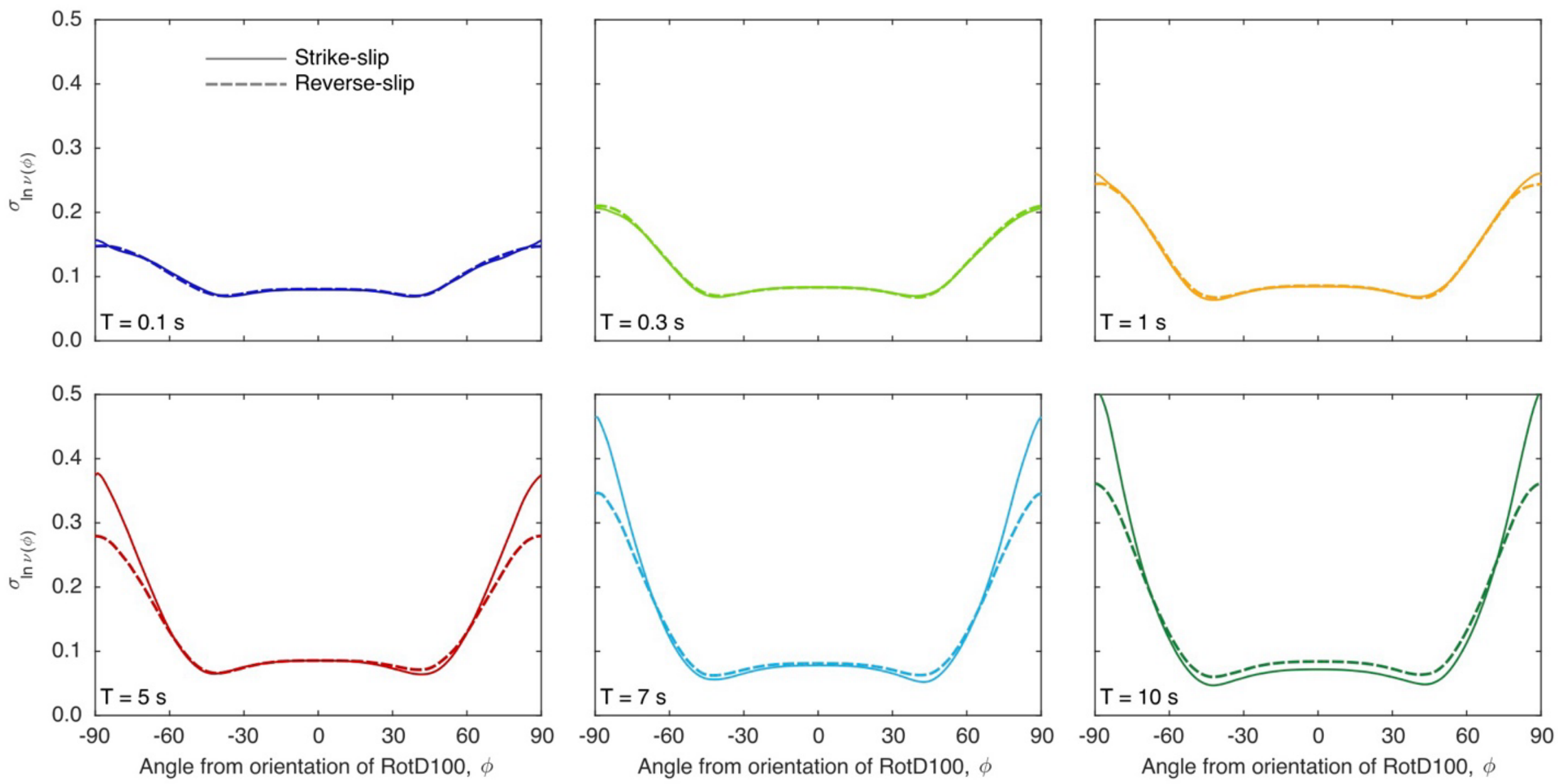

**Figure 4:** Logarithmic standard deviations of $\nu$. Solid lines represent variabilities computed from ground motions recorded in strike-slip earthquakes, and dashed lines represent variabilities computed from ground motions recorded in reverse-slip earthquakes.

***Alt-text:*** *Line graphs showing the logarithmic standard deviations of spectral accelerations at given orientation from RotD100 normalized by RotD50, indicating that standard deviations increase as one rotates away from the major response axis.*

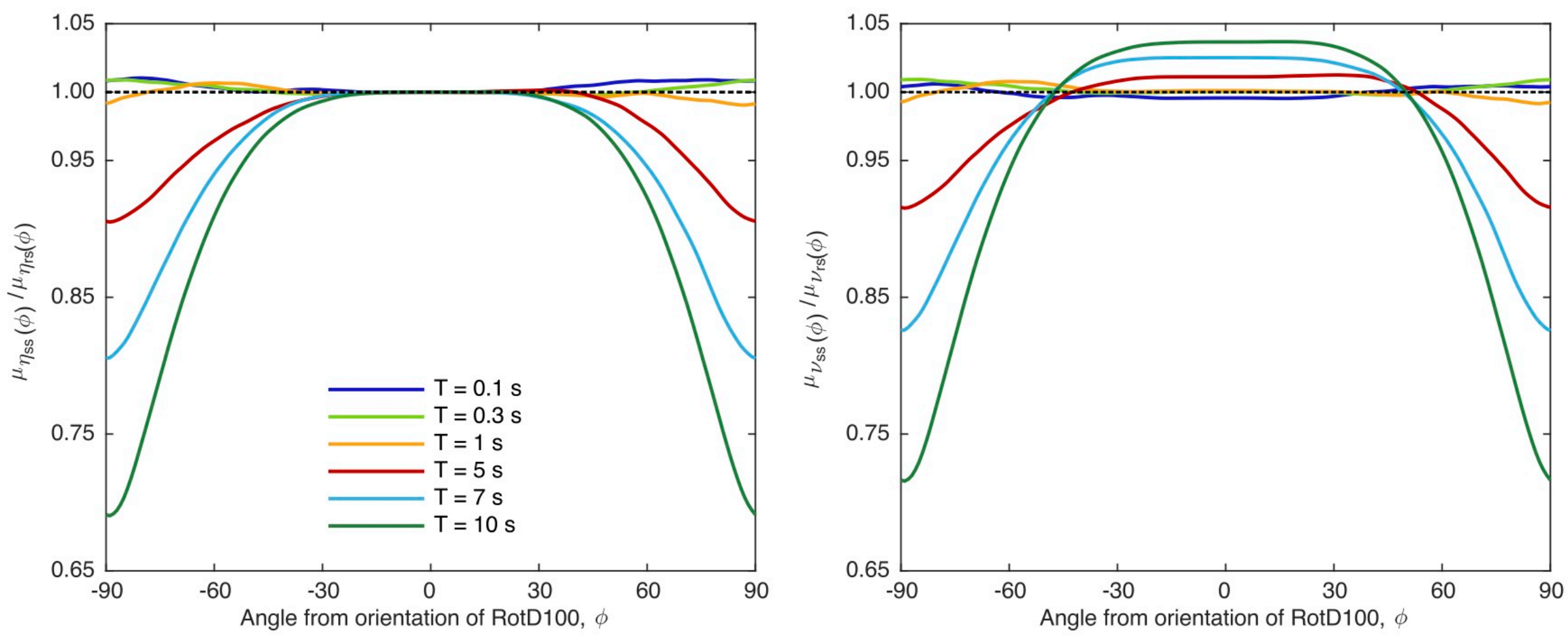

**Figure 5:** Ratios of: (a) geometric mean of $\eta$ for ground motions recorded in strike-slip earthquakes to that from ground motions recorded in reverse-slip earthquakes; (b) geometric mean of $\nu$ for ground motions recorded in strike-slip earthquakes to that from ground motions recorded in reverse-slip earthquakes.

***Alt-text:*** *Line graphs showing ratios of strike-slip earthquakes to reverse-slip earthquakes for the geometric mean of spectral accelerations at given orientation from RotD100 normalized by RotD100 or RotD50, indicating that the ratio decreases as one rotates away from the major response axis.*

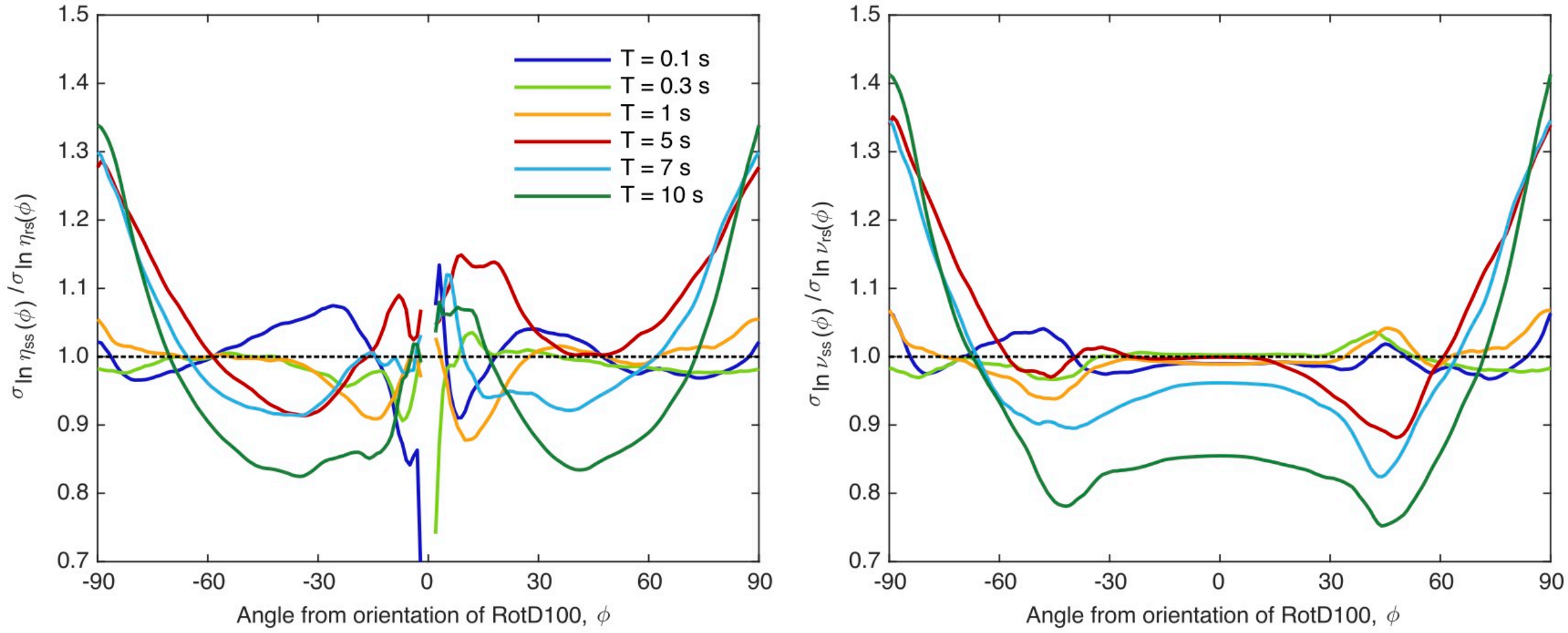

**Figure 6:** Ratios of: (a) logarithmic standard deviations of $\eta$ for strike-slip ground motion records to those for reverse-slip records; (b) logarithmic standard deviations of $\nu$ for strike-slip ground motion records to those for reverse-slip records.

***Alt-text:*** *Line graphs showing ratios of strike-slip earthquakes to reverse-slip earthquakes for the logarithmic standard deviations of spectral accelerations at given orientation from RotD100 normalized by RotD100 or RotD50, indicating that the ratio tends to increase as one rotates away from the major response axis.*

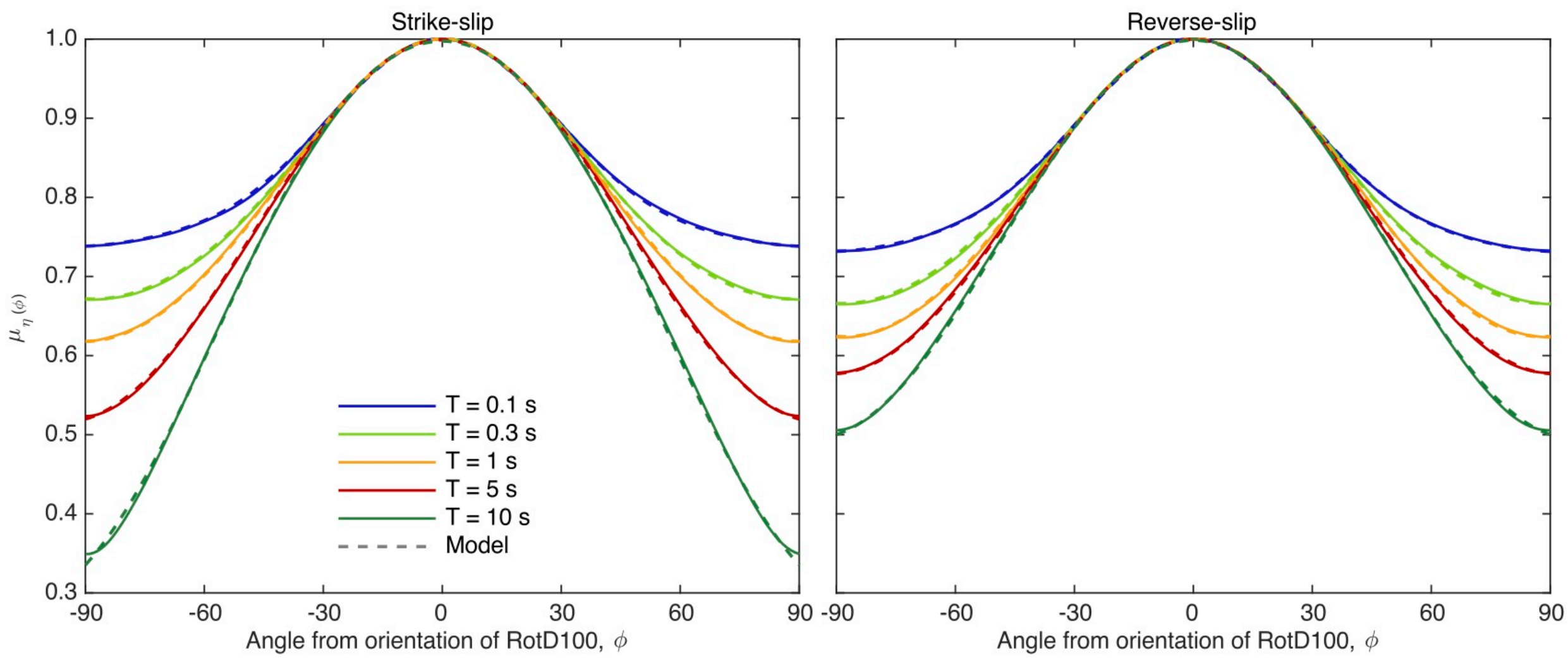

**Figure 7:** Comparison of faulting mechanism specific statistical results and fitted models for the geometric mean of $\eta$. Solid lines indicate empirical results, and dashed lines represent the model (using equation 5).

***Alt-text:*** *Line graphs comparing data and nonlinear regression models for the geometric mean of spectral accelerations at a given orientation from RotD100, normalized by RotD100, indicating that the model lines overlap the data.*

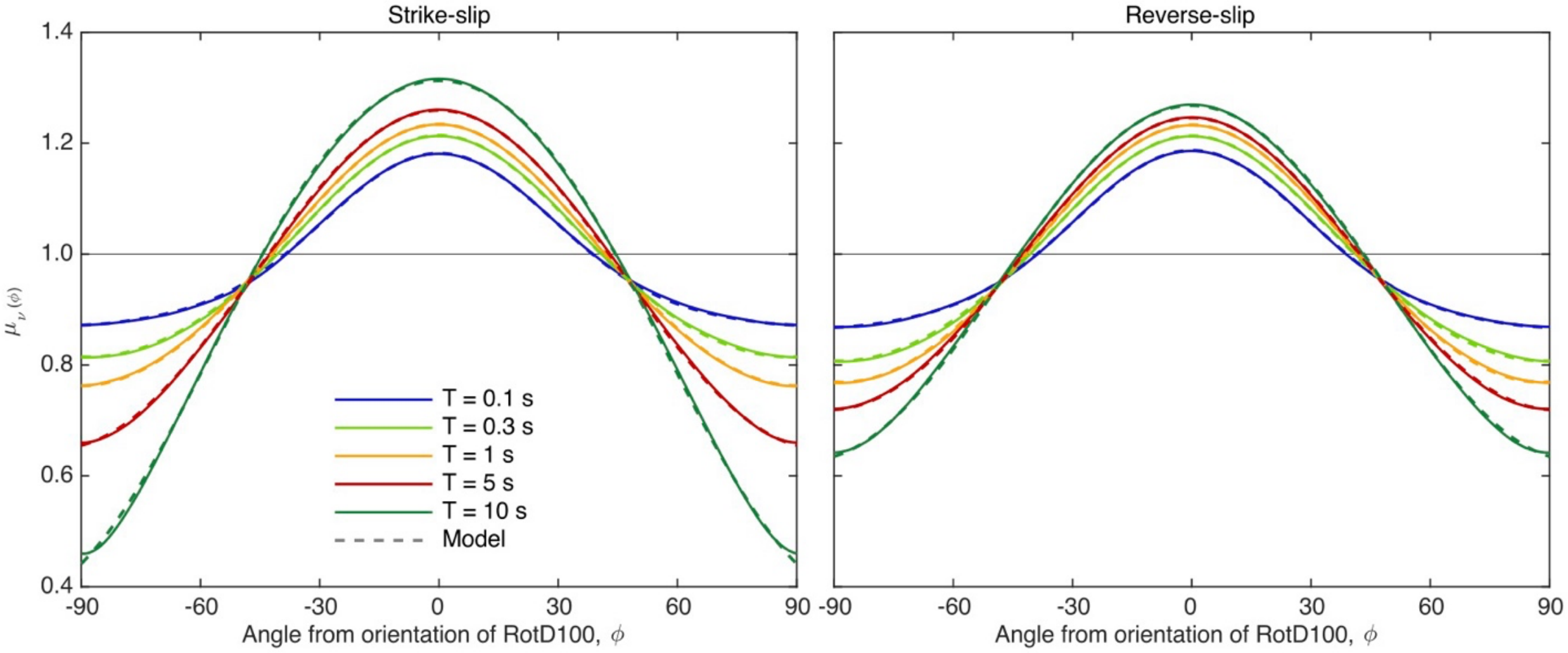

**Figure 8:** Comparison of faulting mechanism specific statistical results and fitted models for the geometric mean of $\nu$ . Solid lines indicate empirical results, and dashed lines represent the model (using equation 5).

***Alt-text:*** *Line graphs comparing data and nonlinear regression models for the geometric mean of spectral accelerations at a given orientation from RotD100, normalized by RotD50, indicating that the model lines overlap the data.*

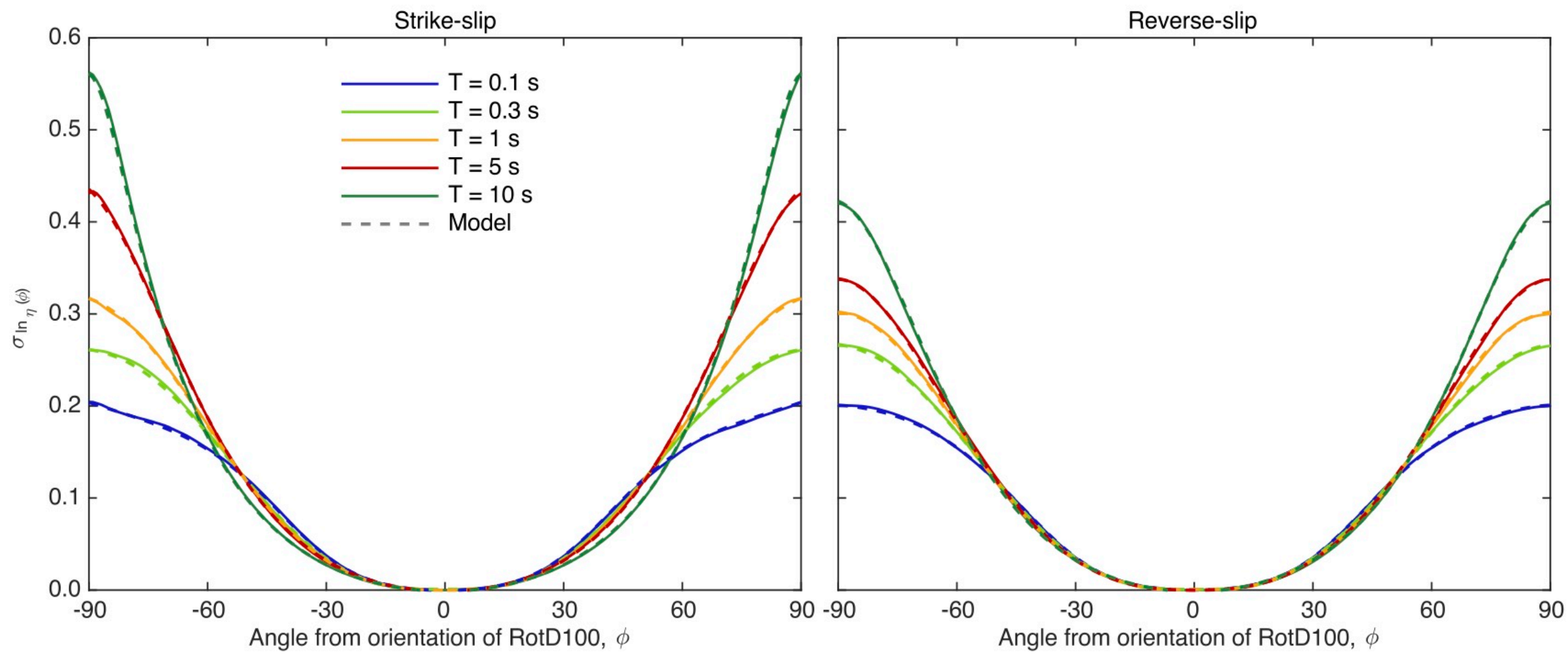

**Figure 9:** Comparison of faulting mechanism specific statistical results and fitted models for the logarithmic standard deviations of $\eta$. Solid lines indicate empirical results, and dashed lines represent the model (using equation 6).

***Alt-text:*** *Line graphs comparing data and nonlinear regression models for the logarithmic standard deviations of spectral accelerations at a given orientation from RotD100, normalized by RotD100, indicating that the model lines overlap the data.*

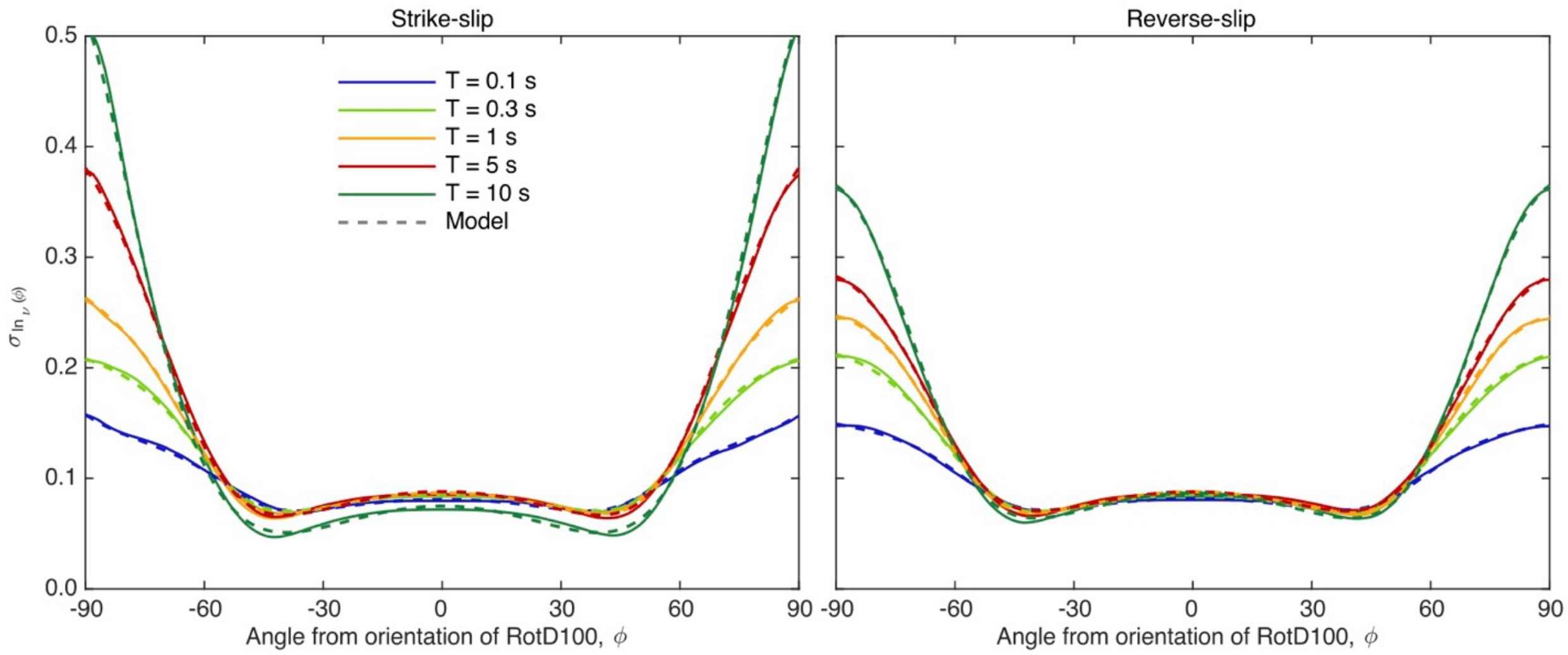

**Figure 10:** Comparison of faulting mechanism specific statistical results and fitted models for the logarithmic standard deviations of $\nu$. Solid lines indicate empirical results, and dashed lines represent the model (using equation 6).

***Alt-text:*** *Line graphs comparing data and nonlinear regression models for the logarithmic standard deviations of spectral accelerations at a given orientation from RotD100, normalized by RotD50, indicating that the model lines overlap the data.*

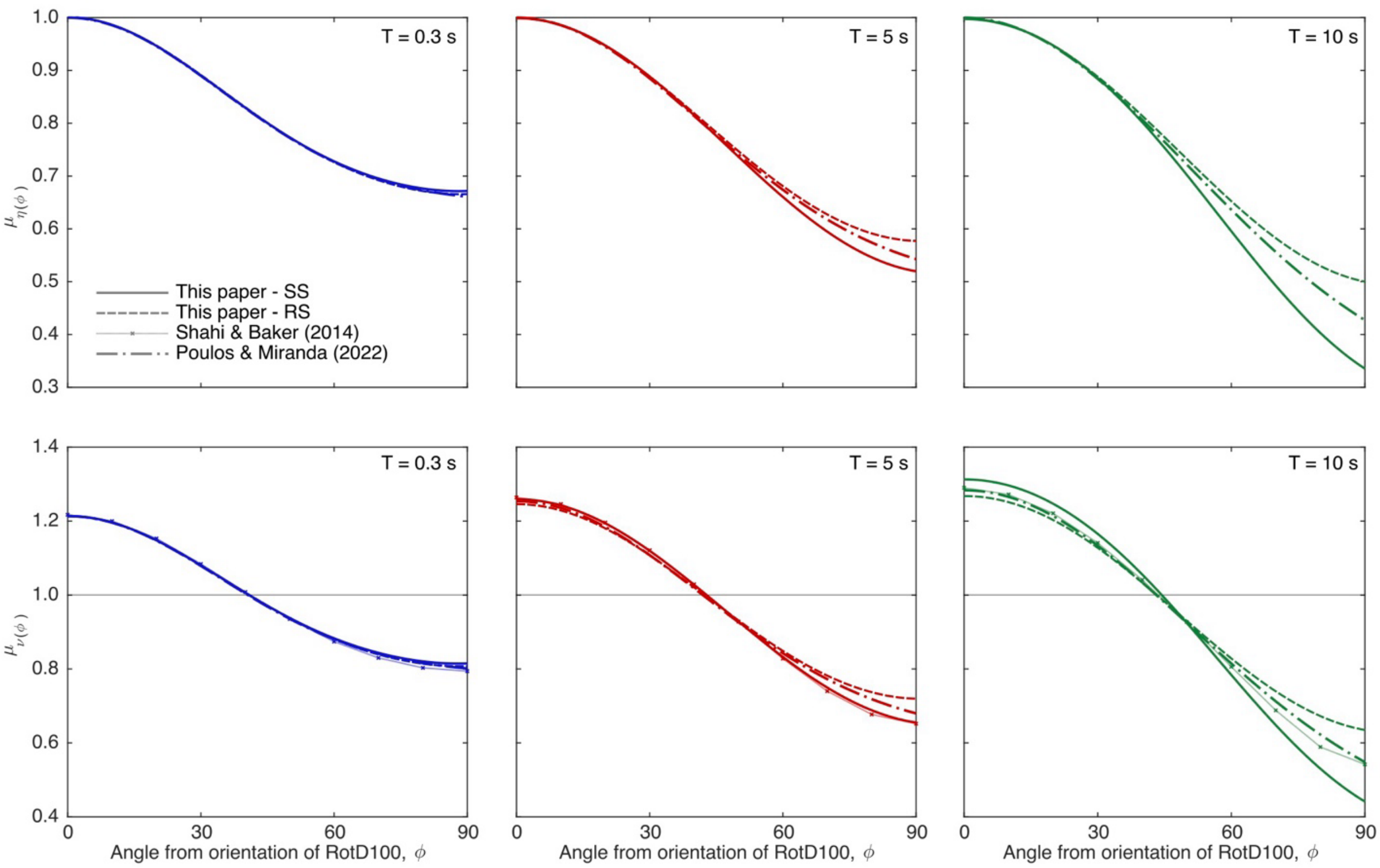

**Figure 11:** Comparison of faulting mechanism specific models for the geometric mean of $\eta$ and $\nu$ (equation 5) from this study with models by Shahi and Baker (2014) and Poulos and Miranda (2022) for crustal earthquakes. The top row presents the comparison for $\eta$ while the bottom shows the comparison for $\nu$.

***Alt-text:*** *Line graphs comparing models developed in this article for the geometric mean of spectral accelerations at a given orientation from RotD100, normalized by RotD100 or RotD50, with models by others, indicating that the model lines overlap at orientations close to RotD100, but deviate from each other at orientations far from RotD100.*

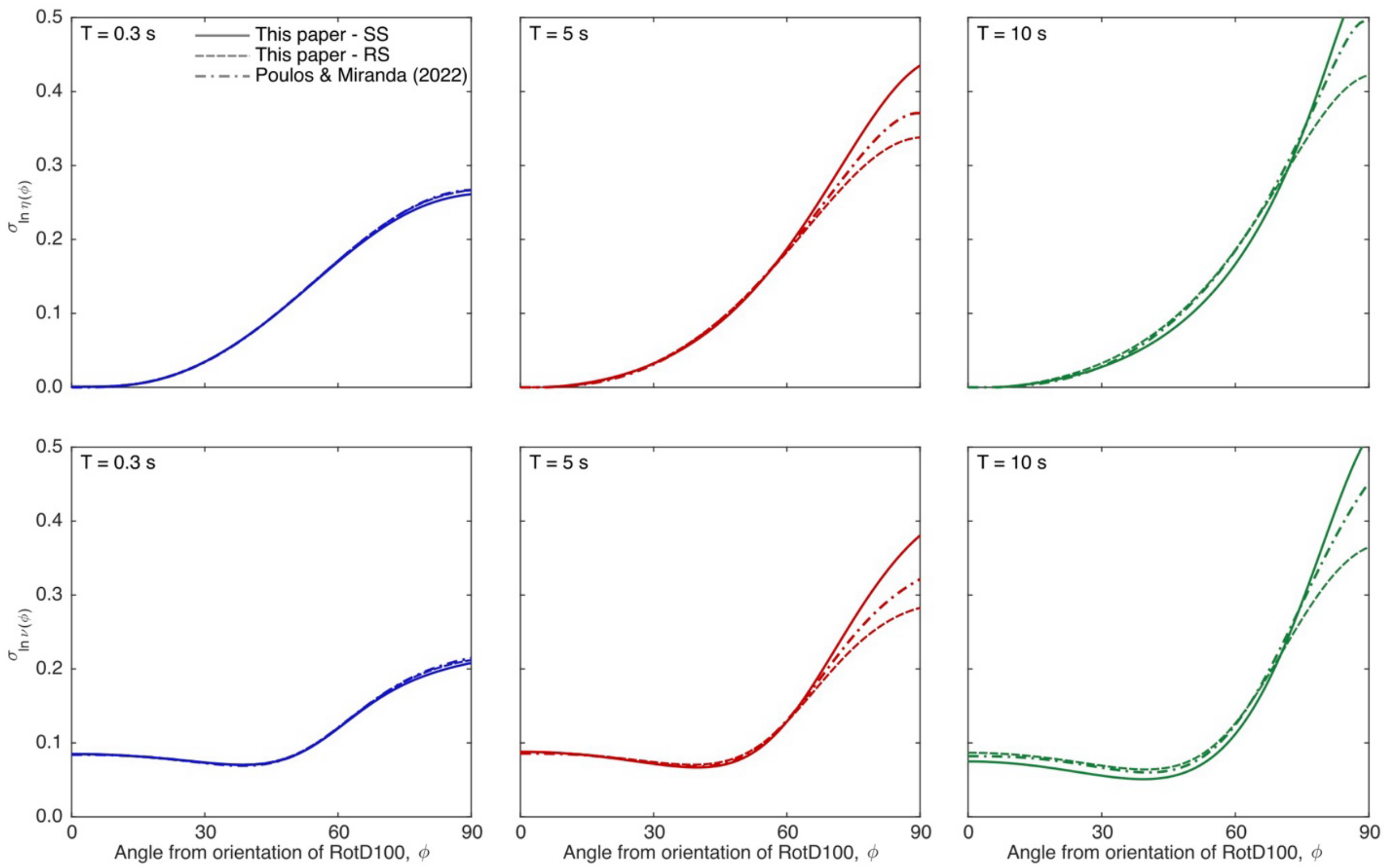

**Figure 12:** Comparison of faulting mechanism-specific models for the logarithmic standard deviations of $\eta$ and $\nu$ developed in this study (equation 6) with the model by Poulos and Miranda (2022). The top row presents the comparison for $\eta$ while the bottom shows the comparison for $\nu$.

***Alt-text:*** *Line graphs comparing models developed in this article for the logarithmic standard deviations of spectral accelerations at a given orientation from RotD100, normalized by RotD100 or RotD50, with models by others, indicating that the model lines overlap at orientations close to RotD100, but deviate from each other at orientations far from RotD100.*

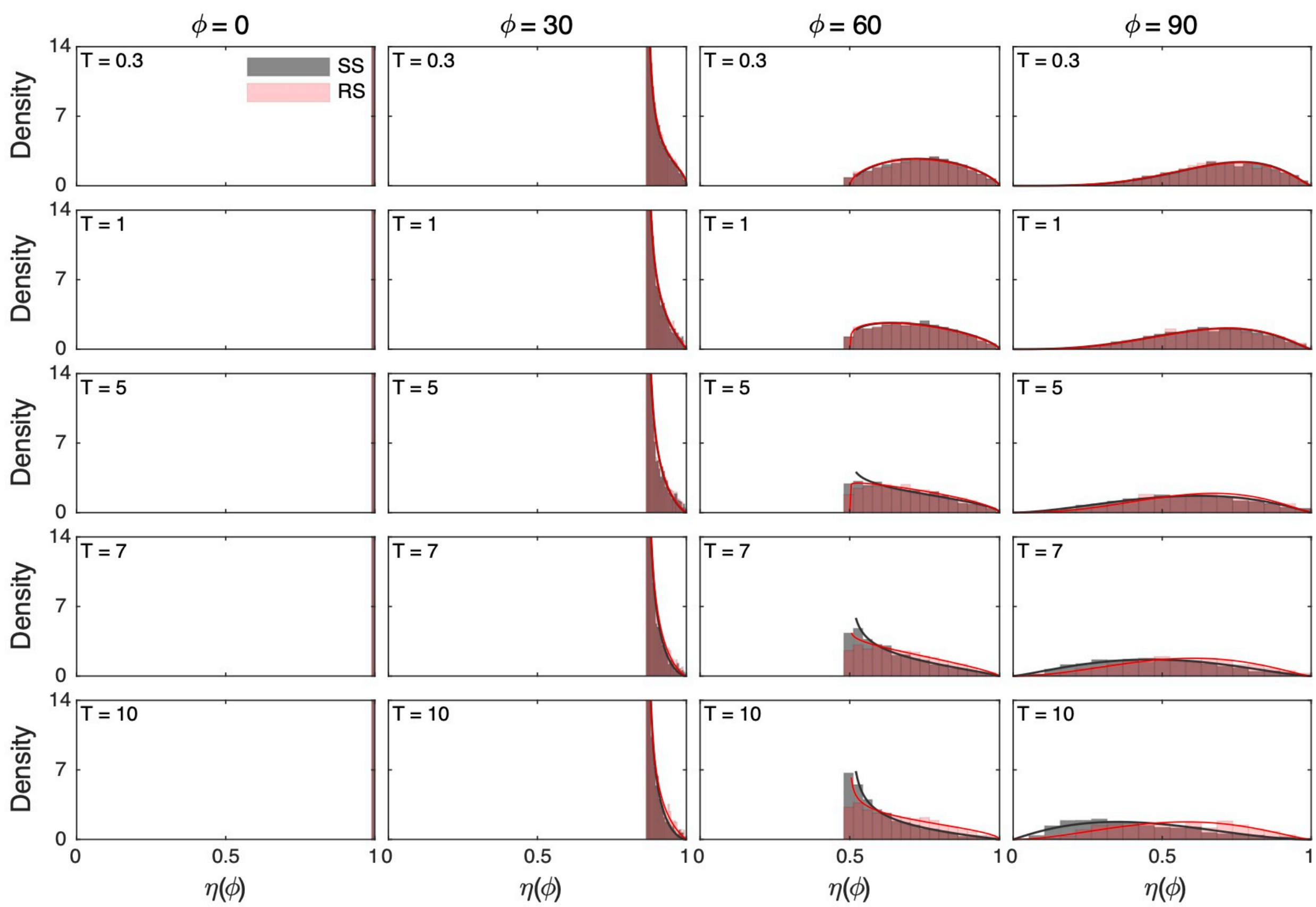

**Figure 13:** Empirical and fitted probability distributions of $\eta$ for strike-slip earthquakes (shown in grey) and reverse-slip earthquakes (shown in red) at four rotation angles and five oscillator periods.

***Alt-text:*** *Histograms showing empirical distributions of spectral accelerations at a given orientation from RotD100, normalized by RotD100, for strike-slip and reverse-slip earthquakes. Overlaid are line graphs showing the fitted distributions, indicating a decent fit to the data.*

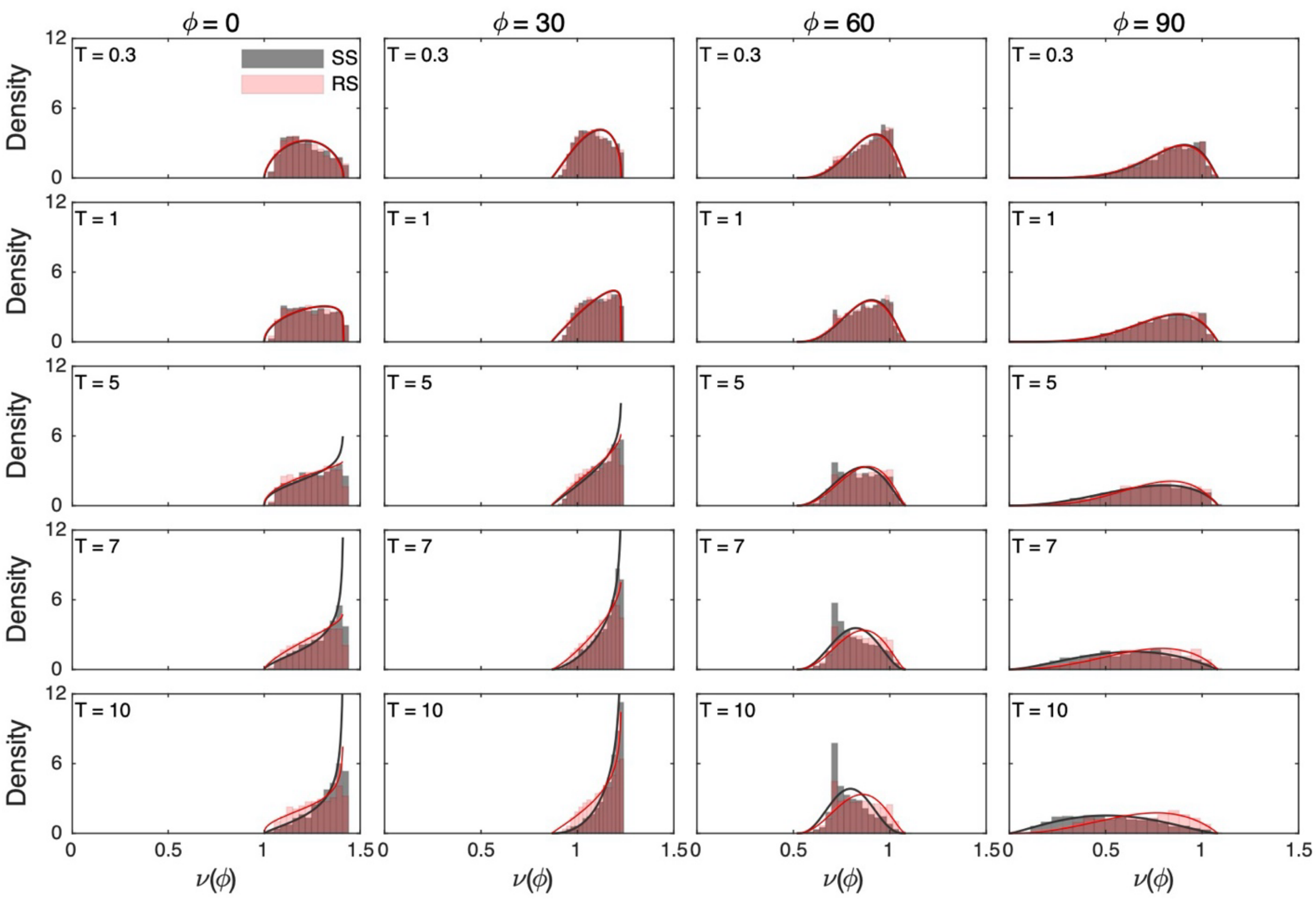

**Figure 14:** Empirical and fitted probability distributions of $\nu$ for strike-slip earthquakes (shown in grey) and reverse-slip earthquakes (shown in red) at four rotation angles and five oscillator periods.

***Alt-text:*** *Histograms showing empirical distributions of spectral accelerations at a given orientation from RotD100, normalized by RotD50, for strike-slip and reverse-slip earthquakes. Overlaid are line graphs showing the fitted distributions, indicating a decent fit to the data.*